\documentclass[aps,prd,preprintnumbers,superscriptaddress]{revtex4}
\usepackage{amsmath}
\usepackage{amssymb}
\usepackage{amsthm}
\usepackage{todonotes}
\usepackage{mathtools}
\usepackage{mathrsfs}
\usepackage{bm}
\usepackage{nameref}
\usepackage{slashed} 
\usepackage{graphicx}
\usepackage{multirow}
\usepackage{tikz}
\usepackage[caption=false]{subfig}
\usepackage{relsize}	
\usepackage{array}
\usepackage{float}
\usepackage{cancel}
\usepackage{color}
\usepackage{xcolor}
\usepackage{soul}
\usepackage{verbatim} 
\usepackage{hyperref}
\usepackage{orcidlink}
\usepackage{pifont}

\allowdisplaybreaks
\newcommand{\beq}{\begin{equation}}
	\newcommand{\eeq}{\end{equation}}
\newcommand{\be}{\begin{eqnarray}}
	\newcommand{\ee}{\end{eqnarray}}

\usepackage[normalem]{ulem}

\begin{document}
	
	\begin{titlepage}
		
		\title{Heavy flavour production in the presence of a weak magnetic field
		}
		
		\author{Sudeep Saha \,\orcidlink{0000-0002-0083-4879}}
		\email{91057@njnu.edu.cn}
		\affiliation{School of Physics and Technology,
			Nanjing Normal University, Nanjing, Jiangsu Province, 210023, China}
        \affiliation{Department of Physical Sciences, Indian Institute of Science Education and Research Berhampur, Berhampur, Odisha 760003, India}
		\author{Sandeep Chatterjee}
		\email{sandeep@iiserbpr.ac.in}
		\affiliation{Department of Physical Sciences, Indian Institute of Science Education and Research Berhampur, Berhampur, Odisha 760003, India}
		\author{Ujjal Kumar Dey}
		\email{ujjal@iiserbpr.ac.in}
		\affiliation{Department of Physical Sciences, Indian Institute of Science Education and Research Berhampur, Berhampur, Odisha 760003, India}
		\author{Snigdha Ghosh}
		\email{snigdha.ghosh@bangla.gov.in}
		\affiliation{Government General Degree College Kharagpur-II, Paschim Medinipur- 721149, West Bengal, India}
		\begin{abstract}
			We calculate analytic expression for the cross section up to order $(eB)^{2}$ for the heavy quark $(Q)$ production of mass $m_{Q}$ through the gluon $(g)$ fusion process $g+g\to Q+\bar{Q}$ in the presence of a constant weak external magnetic field $(B)$ having strength $eB \ll m_{Q}^{2}$. This is done by the weak-field expansion of the fermion propagator calculated using Schwinger’s proper-time formalism. Owing to the hierarchy of scales ($m_Q^2 \gg eB$), we neglect the effects of the magnetic field on the quark and antiquark spinors in the external lines. Since the magnetic field breaks rotational invariance, the cross section becomes anisotropic and depends on the relative orientation angle $(\theta)$ between the magnetic field and the incoming gluons in the center of mass frame. The lowest-order magnetic correction to the vacuum cross section scales as $(eB)^2$ and is found to have $\cos 2\theta$ and $\cos 4\theta$ dependence, reflecting the breaking of rotational symmetry by the magnetic field. The cross section is maximized when the gluons propagate along the field direction ($\theta = 0,~\pi$) and minimized in the perpendicular configuration ($\theta = \pi/2$). The resulting asymmetry peaks near the production threshold and falls with increasing center of mass energy, scaling as $(eB)^2/m_Q^4$ with the quark mass. These results provide direct, calculable signatures of the initial-stage magnetic field in relativistic heavy-ion collisions which can, in principle, be imprinted on heavy-flavour observables measured in the final state of heavy-ion collisions.
		\end{abstract}
		\date{\today}
		\maketitle
		
	\end{titlepage}
	\section{Introduction}
    Heavy-ion collision (HIC) experiments serve as the primary tool for investigating the properties of nuclear matter under extreme conditions of energy density. By accelerating heavy nuclei such as gold and lead to ultra-relativistic speeds at facilities like the Relativistic Heavy Ion Collider (RHIC) and the Large Hadron Collider (LHC), these collisions aim to recreate the quark–gluon plasma (QGP) \cite{Shuryak:1980tp}, a deconfined state of matter where quarks and gluons are no longer bound within hadrons, thereby reproducing conditions analogous to those of the early universe microseconds after the Big Bang. \cite{Kolb:1990vq}. The resulting experimental data provide essential constraints on medium properties of the strongly interacting matter governed by the rules of Quantum Chromodynamics (QCD), particularly regarding the phase transition between hadronic matter and QGP, the restoration of chiral symmetry, and the manifestation of P and CP violations in strong interactions \cite{Ko2023, Chen:2024aom, Shou:2024uga, Liu:2020ymh}. 

    In non-central HICs, the relativistic motion of charged spectator protons generates intense magnetic fields, reaching strengths of approximately $10^{18}~\text{G}$ at RHIC and $10^{19}~\text{G}$ at LHC \cite{Skokov:2009qp, Kharzeev:2007jp, Deng:2012pc}. These field strengths approach the scale of the strong interaction itself, making electromagnetic contributions to QGP dynamics non-negligible \cite{Yan:2021zjc}. A number of studies indicate that the strong magnetic fields produced in non-central HICs give rise to several unique and intriguing properties. The electromagnetic field provides us with a chance to search for novel transport phenomena like the chiral magnetic effect \cite{Fukushima:2008xe, Basar:2012gm, Kharzeev:2007jp} and chiral vortical effect \cite{Kharzeev:2007tn, Kharzeev:2010gr}, photoproduction of dileptons and quarkonia in peripheral and ultra-peripheral collisions \cite{Zha:2018ytv, ALICE:2013wjo, ALICE:2015mzu, STAR:2018ldd, STAR:2019wlg}, splitting of $D^{0}$ and $\overline{D}^{0}$ directed flows \cite{Das:2016cwd, STAR:2019clv, ALICE:2019sgg} and spin-polarized difference between $\Lambda$ and $\overline{\Lambda}$ \cite{Muller:2018ibh, Guo:2019joy}.

    The initial strength of the magnetic field created in HICs has been calculated using classical electrodynamics (like employing Biot-Savart law) in Refs. \cite{Skokov:2009qp, Kharzeev:2007jp, Deng:2012pc, Zhao:2019crj, Zhao:2022dac}. However, how this magnetic field decays over time as the spectators fly away from the mid-rapidity region is still an open question. The difficulty comes from the complex interaction between the magnetic field and the charge-conducting medium \cite{Yan:2021zjc, Tuchin:2015oka, Huang:2022qdn, Li:2023tsf, Wang:2021oqq, Zhang:2022lje, Huang:2024aob}. This interaction is described by Maxwell's equations and depends strongly on the temperature-dependent electrical conductivity of the medium \cite{Ding:2010ga, Francis:2011bt, Amato:2013naa}. The properties of the medium before it reaches equilibrium also affect the magnetic field's strength as it transits to a hydrodynamic evolution \cite{Yan:2021zjc}. Moreover, the medium may not always follow Ohm's law, which could further influence the field's time evolution \cite{Wang:2021oqq}. Recent research also suggests that the vorticity of the QGP fluid might prolong the magnetic field's lifetime \cite{Huang:2024aob, Guo:2019mgh}. Because of all these complications, different theoretical models predict magnetic field strengths that can differ by several orders of magnitude \cite{Yan:2021zjc, Li:2023tsf, Wang:2021oqq, Huang:2024aob, McLerran:2013hla, Voronyuk:2011jd, Stewart:2021mjz}. Due to these theoretical uncertainties, there is a need for experimental investigations of the magnetic field in HICs, along with more theoretical investigations, to better understand the effect of the electromagnetic field produced in the HIC.
    
    As the electromagnetic field is strongest at the initial stages of the collision, heavy-flavour hadrons are arguably one of the best probes to detect their presence for several reasons \cite{Das:2016cwd,Zhao:2020jqu}.
    Firstly, the masses of heavy quarks are much larger than the temperature of the QGP produced at RHIC and LHC energies ($m_{Q} \sim (1-5)~\text{GeV}\gg T \sim (0.3-0.5)~\text{GeV}$) \cite{Zhang:2007yoa, Andronic:2015wma}. As a result, they are mainly produced in the initial hard collisions and not thermally generated within the QGP. Secondly, the time scale for heavy quark formation ($\Delta \tau \sim 1/2m_{Q} \sim 0.07~\text{fm}$ for charm quark and $0.02~\text{fm}$ for bottom quark) is much shorter than the QGP formation time ($\tau_{0} \sim 0.5~ \text{fm}$) at RHIC and LHC energies. This means that once produced, heavy quarks experience the entire space–time evolution of the fireball. They lose some of their energy by interacting with the QGP constituents and also take part in the collective motion of the system. 
    Additionally, since heavy flavours are produced almost entirely at the beginning of the collision, they should be significantly influenced by the electromagnetic field generated in HICs. As a result, their modified properties will be passed on to the heavy-flavour hadrons detected in the final state, making them excellent probes. 
    
    Motivated by this, in the present work we intend to calculate the cross section for heavy quark production via the gluon fusion process, $g + g \rightarrow Q + \bar{Q}$, in the presence of an external magnetic field. The magnetic field in high-energy proton–proton collisions was first estimated in Ref. \cite{Ambjorn:1990jg}, and for heavy-ion collisions in Ref. \cite{Kharzeev:2007jp}, by using a realistic proton distribution within the nucleus. The electromagnetic field in vacuum for a heavy-ion collision can be obtained using Liénard–Wiechert potentials, combined with standard models of nuclear charge density to locate the protons inside the heavy ions \cite{Tuchin:2013ie, Tuchin:2012mf}. Numerical calculations show that the magnetic field is very strong immediately after the collision but decays rapidly, roughly as the inverse square of time \cite{Kharzeev:2007jp, Fukushima:2008xe, Mustafa:2025uad}. The presence of a medium in the collision slows down this decay. The relaxation time of the magnetic field is longer in an expanding medium (where the temperature changes with time) than in a static medium (where the temperature is constant) \cite{Tuchin:2013ie, Tuchin:2013apa, Tuchin:2014hza, Mustafa:2025uad}. For simplicity, we do not include the effects of the medium in this work; this is left for future studies. Since the magnetic field in vacuum decays quickly, we treat it as weak and also constant for simplicity in our calculation of the gluon fusion process. Additionally, the masses of the heavy quarks produced are much larger than the typical magnetic field scale ($m^2_Q \gg eB$, where $eB\sim m_\pi^2\sim 10^{18}~\text{G}$), which further justifies our weak-field approximation. The magnetic field enters the process through the internal fermion propagator and the external lines for the heavy quark pair. We use Schwinger's proper-time method \cite{Schwinger:1951nm} to determine the expression of the fermion propagator in an external weak magnetic field, and since $\sqrt{eB}\ll m_{Q}$, we ignore the effect of the magnetic field on the external fermion lines \cite{Bhattacharya:2002qf}. As the external magnetic field breaks the rotational invariance, the cross-section becomes anisotropic and explicitly dependent on the relative orientation angle $(\theta)$ between the magnetic field and the incoming gluons momenta
    
    The remainder of this paper is organised as follows. In Sec.~\ref{sec:propagator}, we introduce the Schwinger proper-time formalism and derive the expression for the fermion propagator in the presence of a constant background magnetic field $\boldsymbol{B}=B\hat{z}$, adopting the Landau gauge. The full propagator is then expanded perturbatively in powers of $eB$ in the weak-field limit $eB \ll m_Q^2$. Next, in Sec.~\ref{sec:crosssection} along with its three subsections, we systematically calculate the heavy quark production cross section for the gluon fusion process $g+g\to Q+\bar{Q}$ in the presence of a weak external magnetic field; the calculation of the invariant amlitudes in shown in subsection \ref{Amplitude squared of different channels} followed by the discussions on the kinematics in subsection \ref{Kinematic configuration in the presence of a magnetic field}. In subsection \ref{Cross-section for the process}, we obtain the analytic expression of the total cross-section up to order $(eB)^2$, supplemented by a detailed numerical analysis in Sec. \ref{Numerical analysis}. Finally, Sec.~\ref{sec:conclusion} contains a summary of the main results and our conclusions. Some of the calculational details are provided in the appendix \ref{Explicit expressions for the amplitude squared in a weak magnetic field}.
\section{Fermion Propagator in a Weak Background Magnetic Field}
\label{sec:propagator}
For the process $g+g\to Q+\bar{Q}$ in heavy-ion collisions, we assume that the produced heavy quarks (charm or bottom) with mass $m_{Q}$ satisfy $m^2_Q \gg eB$, where $eB\sim m_\pi^2\sim 10^{18}~\text{G}$ is the typical magnetic field strength generated in non-central collisions \cite{Kharzeev:2007jp,Skokov:2009qp,Voronyuk:2011jd,Deng:2012pc,Tuchin:2013ie}. This scale hierarchy justifies a weak-field expansion of the fermion propagator, treating the magnetic field as a perturbative correction to the free Dirac propagator.  
	
The Dirac equation for a fermion of mass $m_{Q}$ and electric charge $e$ in a classical background electromagnetic field $A_{\mu}(x)$ is 
\begin{align}
	\left(i \gamma^{\mu}D_{\mu}-m_{Q} \right) \psi(x) = 0, 	\end{align}
where $\gamma^\mu$ are the Dirac matrices satisfying $\{\gamma^\mu,\gamma^\nu\}=2g^{\mu\nu}$, $D_{\mu} = \partial_{\mu}+ i e A_{\mu}(x)$ and we have used the metric tensor with signature $g^{\mu\nu}=\text{diag}\left(1,-1,-1,-1\right)$. For a constant magnetic field $\boldsymbol{B}=B\hat{z}$ along the positive $z$-direction, we adopt a symmetric gauge $A_\mu = \frac{B}{2}(0, -y, x, 0), \label{landau_gauge}$ which yields the field strength tensor \(F_{12}=-F_{21}=B\), with all other components vanishing. This gauge preserves the momentum along the $z$-axis \cite{Bhattacharya:2007vz}. The magnetic field breaks the rotational symmetry, as a result, rotation about the field axis (i.e. the $z$-axis) remains a symmetry $-$ introducing anisotropy between the directions parallel and perpendicular to the magnetic field. To facilitate the calculation, it is customary to decompose the four-vector components into parallel \((\parallel)\) (in the direction of the magnetic field) and perpendicular \((\perp)\) components:
	\begin{align}
		a^{\mu} &= (a^{0}, a^{1}, a^{2}, a^{3}) = (a_{0}, \bm{a}); \qquad 
		a \cdot b \equiv a_{0}b_{0} - \bm{a} \cdot \bm{b}; \qquad 
		g^{\mu\nu} = \text{diag}(1, -1, -1, -1), \notag \\
		a^{\mu} &= a_{\parallel}^{\mu} + a_{\perp}^{\mu}; \qquad 
		a_{\parallel}^{\mu} = (a^{0}, 0, 0, a^{3}); \qquad 
		a_{\perp}^{\mu} = (0, a^{1}, a^{2}, 0), \notag \\
		g^{\mu\nu} &= g_{\parallel}^{\mu\nu} + g_{\perp}^{\mu\nu}; \qquad 
		g_{\parallel}^{\mu\nu} = \operatorname{diag}(1, 0, 0, -1); \qquad 
		g_{\perp}^{\mu\nu} = \operatorname{diag}(0, -1, -1, 0), \notag \\
		(a \cdot b) &= (a \cdot b)_{\parallel} - (a \cdot b)_{\perp}; \qquad 
		(a \cdot b)_{\parallel} = a^{0}b^{0} - a^{3}b^{3}; \qquad 
		(a \cdot b)_{\perp} = a^{1}b^{1} + a^{2}b^{2}, \notag \\
		\slashed{a} &= \gamma^{\mu}a_{\mu} = \slashed{a}_{\parallel} - \slashed{a}_{\perp}; \qquad 
		\slashed{a}_{\parallel} = \gamma^{0}a_{0} - \gamma^{3}a^{3}; \qquad 
		\slashed{a}_{\perp} = \gamma^{1}a^{1} + \gamma^{2}a^{2}. \label{convention}
	\end{align}
	
	\subsection{Schwinger proper-time representation}
	
	The fermion propagator in a constant background field admits an elegant representation via Schwinger's proper-time method \cite{Schwinger:1951nm}. Working in the operator formalism, we denote \(x_{\mu}\) and \(p_{\mu}\) as the eigenvalues of the position and the momentum operators, respectively. The corresposnding operators in the Hilbert space is denoted by \(X_{\mu}\) and \(P_{\mu}\) respectively:
	
	\begin{align}
		P_{\mu} = i \partial_{\mu}, \quad X_{\mu} |x\rangle = x_{\mu} |x\rangle
	\end{align} 
	The covariant momentum is \(\Pi_\mu = P_\mu -  e A_\mu(X)\), satisfying \([\Pi_\mu,\Pi_\nu]=-i e F_{\mu\nu}\). 
	In Schwinger's proper time method, the propagator can be expressed as 

\begin{align}
	S(x,x^{\prime}) = -i \int_{-\infty}^{0} ds~ U(x,x^{\prime},s),
\end{align}
	where \(U(x,x^{\prime},s)\) is defined as 
	\begin{align}
		U(x,x^{\prime},s) = \langle x |e^{-isH_{f}}| x^{\prime} \rangle, 
	\end{align}
	which satisfies 
	\begin{align}
		i\frac{\partial U}{\partial s} = H_{f} U, \quad U(x,x^{\prime};0) = \delta^{(4)}(x-x^{\prime}).
	\end{align}
	The fermionic Hamiltonian is defined as \(H_{f}= -\left(\gamma \cdot \Pi\right)^{2}\) and the propagator satisfies 
	\begin{align}
		\left(\gamma \cdot \Pi-m_{Q}\right)S(x,x^{\prime}) = \delta^{(4)}\left(x-x^{\prime}\right).
\end{align}	
Using Schwinger's proper time method, the propagator can be expressed as
\begin{align}
	S(x,x') = \Phi^{\prime}(x,x')~\tilde{S}\left(x-x^{\prime}\right),
\end{align}
where \(\Phi^{\prime}(x,x')=i (4\pi)^{2}\exp\left(i e \int_{x^{\prime}}^{x}d\xi^{\mu}\left[A_{\mu}(\xi)+\frac{1}{2}F_{\mu\nu}(\xi-x^{\prime})^{\nu}\right]\right)\) is the gauge-dependent Schwinger phase factor and $\tilde{S}\left(x-x^{\prime}\right)$ is the translationally invariant part. The phase factor is independent of the chosen path. Thus, by taking a straight-line path along with choosing an appropriate gauge transformation, we can gauge away the phase factor when a single fermion propagator is involved \cite{Ayala:2014uua}. In the proper-time formalism, the translationally invariant part takes the following form
\begin{align}
\nonumber    \tilde{S}\left(x-x^{\prime}\right) =& -\frac{1}{(4\pi)^{2}} \int_{0}^{\infty}ds \frac{e^{-im_{Q}^{2}s}+ieBs\Sigma_{3}}{s \sin{eBs}} ~\exp\left[-\frac{i}{4s}\left((x-x^{\prime})_{\parallel}^{2}-\frac{eBs}{\tan{eBs}}(x-x^{\prime})_{\perp}^{2}\right)\right] \\& \times \left[m_{Q}+\frac{1}{2s}\left(\gamma \cdot (x-x^{\prime})_{\parallel}-\frac{eBs}{\sin{eBs}}e^{-ieBs\Sigma_{3}}\gamma \cdot (x-x^{\prime})_{\perp}\right)\right],
\end{align}
where \(\Sigma_3 = i\gamma^1\gamma^2\) is the spin operator.
The translationally invariant part $\tilde{S}\left(x-x^{\prime}\right)$, upon Fourier transformation, takes the following form in momentum space \cite{Mustafa:2025uad,Schwinger:1951nm,Chyi:1999fc}:

\begin{align}
	iS(p) &= \int_{0}^{\infty} ds \, \frac{1}{\cos(eBs)} \, 
	\exp\left\{ i s\left[ p_{\parallel}^{2} - \frac{\tan(eBs)}{eBs} \, p_{\perp}^{2} - m_{Q}^{2} \right] \right\} 
	\notag \\
	&\quad \times \left[ e^{i eBs \Sigma_{3}} (m_{Q} + \gamma \cdot p_{\parallel}) - 
	\frac{\gamma \cdot p_{\perp}}{\cos(eBs)} \right].
	\label{fermi_prop}
\end{align}
	
	 This is the Schwinger proper time representation of the fermion propagator in a constant magnetic field.
	
\subsection{Weak-field approximation}
We consider the weak-field approximation to the heavy quark production via the gluon fusion process, since the heavy quarks produced satisfy \(\sqrt{eB}\ll m_{Q}\). In this case, we can treat it as a perturbation and expand the fermion propagator in Eq. \ref{fermi_prop} in powers of the magnetic field \(eB\). We can write the propagator up to the order of \((eB)^2\) as \cite{Mustafa:2025uad,Chyi:1999fc,Ayala:2004dx,Haque:2017nxq,Bandyopadhyay:2017cle}:
\begin{align}
	S_{w}(p) = S_0(p) + S_1(p) + S_2(p)
\end{align}
with
\begin{align}
	S_0(p) =& \frac{\slashed{p}+m_{Q}}{p^2-m_{Q}^2},\label{LOpropagator} \\
	S_1(p) =& e B\,\frac{i(\slashed{p}_\parallel+m_{Q})\gamma^1\gamma^2}{(p^2-m_{Q}^2)^2},\label{correctionPropagator} \\
	S_2(p) =& -2(e B)^2\left[\frac{\slashed{p}_\perp}{(p^2-m_{Q}^2)^3} - \frac{p_\perp^2(\slashed{p}+m_{Q})}{(p^2-m_{Q}^2)^4}\right].
\end{align}
Although we expand the propagator to $(eB)^{2}$, the leading magnetic field correction to the cross section arises through $S_{1}(p)$. The contribution of $S_{2}(p)$ to the amplitude involves additional powers of $eB/m_{Q}^{2}\ll 1$ relative to $S_{1}(p)$ and is therefore neglected in this present work.
The first term \(S_0(p)\) is the standard free Dirac propagator. The term linear in \(e B\), namely \(S_1(p)\), explicitly breaks rotational invariance by involving only the parallel momentum components. Apart from that, since the magnitude of the magnetic field is much smaller than the mass of the heavy quark produced, we ignore its effects on the quark and antiquark spinors \cite{Bhattacharya:2002qf}.

\section{Heavy quark production cross-section in a magnetic field}
\label{sec:crosssection}
In relativistic heavy-ion collisions, heavy quarks are predominantly produced through initial hard partonic scatterings. At leaing order, the relevant production channels are the gluon fusion $g+g\to Q+\bar{Q}$ and quark anti-quark anniilation $q+\bar{q}\to Q+\bar{Q}$ \cite{Nason:1987xz}. For the nucleus-nucleus collision centre-of-mass energy ($\sqrt{s_{NN}}$) at RHIC and LHC, the gluon fusion process is far more dominant mechanism. This dominance arises from the fact that the parton distribution functions of nuclei at small Bjorken-$x$ (where $x \sim 2m_{Q}/\sqrt{s_{NN}}\sim 10^{-3} - 10^{-2}$ for charm and bottom production) are overwhelmingly populated by gluons, while the quark and antiquark densities are comparatively suppressed \cite{Chen:2024lmp}. This is also confirmed by the NNPDF4.0 \cite{NNPDF:2021njg} determination, that the gluon parton distribution functions completely dominates over all quark and antiquark parton distribution functions at small-$x$. This dominance means that processes initiated by gluon-gluon fusion have much larger cross sections than those driven by quark-antiquark annihilation.
\subsection{Amplitude squared of different channels} \label{Amplitude squared of different channels}
\begin{figure}[h]
	\includegraphics[scale=0.35]{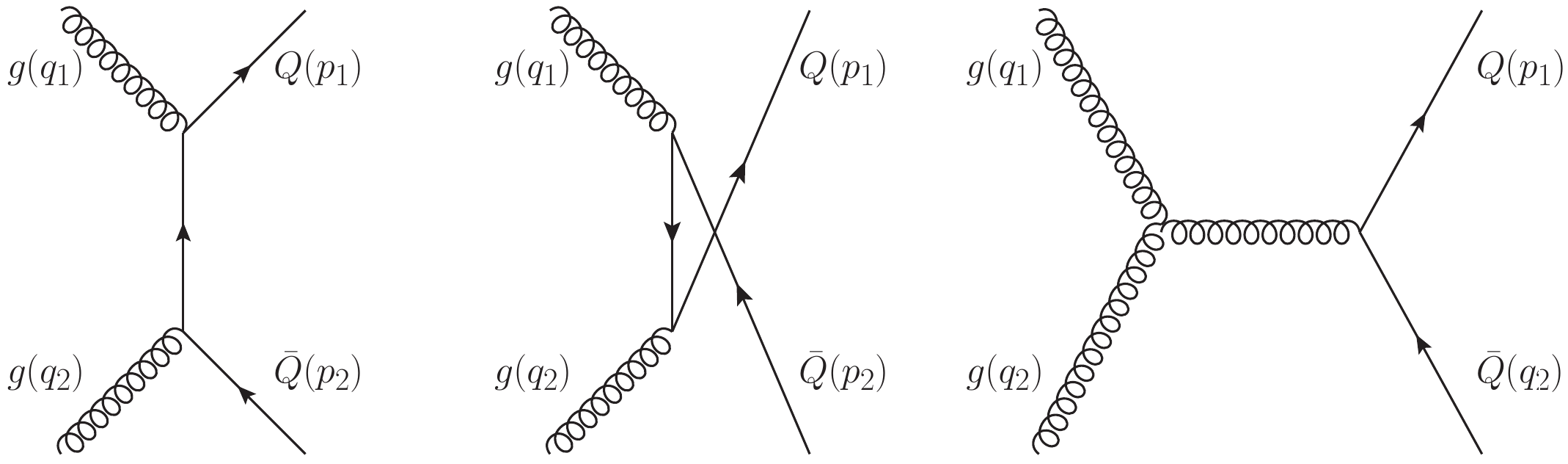}
	\caption{The $t$, $u$ and $s$ channels of heavy quark production process \(g (q_{1})+g(q_{2}) \to Q(p_{1})+\bar{Q}(p_{2})\) at leading order.}
	\label{feynman diagrams}
\end{figure}

We define the Mandelstam variables as 
\begin{align}
	s = \left( q_{1}+q_{2} \right)^{2} = \left(p_{1}+p_{2}\right)^{2}, \quad
	t = \left(q_{1}-p_{1}\right)^{2} = \left(q_{2}-p_{2}\right)^{2}, \quad
	u = \left(q_{1}-p_{2}\right)^{2} = \left(q_{2}-p_{1}\right)^{2},
\end{align}
which satisfy the relation $s+t+u = 2m_{Q}^{2}$. At leading order, the \(t, u\) and \(s\) channels of the process \(g (q_{1})+g(q_{2}) \to Q(p_{1})+\bar{Q}(p_{2})\) are shown in Fig. \ref{feynman diagrams}. The total amplitude squared of different channels can be written as 
\begin{align}
    |\mathcal{M}|^2 = |\mathcal{M}_{t}+\mathcal{M}_{u}+\mathcal{M}_{s}|^{2} = |\mathcal{M}_{t}|^2+|\mathcal{M}_{u}|^2+|\mathcal{M}_{s}|^2+2\text{Re}|\mathcal{M}_{t}\mathcal{M}^{\dagger}_{u}+\mathcal{M}_{t}\mathcal{M}^{\dagger}_{s}+\mathcal{M}_{u}\mathcal{M}^{\dagger}_{s}|.
    \label{totalamplitudesquared}
\end{align}
We first consider the \(t-\)channel, the amplitude squared for this channel is given by 
\begin{align}
\left | \mathcal{M}_{t} \right |^2 = g_{s}^{4} \mathcal{C}_{tt} \text{Tr}\left [\gamma_{\mu}\left ( \slashed{p}_{1}+m_{Q} \right) \gamma^{\mu} S_{w}(p_{1}-q_{1}) \gamma^{\nu} \left( \slashed{p}_{2} -m_{Q}\right) \gamma_{\nu} S_{w}^{\dagger}(p_{1}-q_{1})    \right ],
\end{align}
where the colour factor is \(\mathcal{C}_{tt} = \sum_{\text{col}} \text{Tr} \left [t^{a}t^{b}t^{b}t^{a} \right ] = C_{A}C_{F}^{2} = \frac{16}{3}\), $g_{s}$ is the gauge coupling constant of the $SU(3)$ colour group, and we have taken the fermion propagator up to order \(eB\), i.e.
\begin{align}
 S_{w}(p) =  \frac{\slashed{p}+m_{Q}}{p^2-m_{Q}^2} + e B\,\frac{i(\slashed{p}_\parallel+m_{Q})\gamma^1\gamma^2}{(p^2-m_{Q}^2)^2}. \label{propagator}
\end{align}
After expanding the amplitude squared, it can be written as
\begin{align}
	\left | \mathcal{M}_{t} \right |^2 = g_{s}^{4} \mathcal{C}_{tt} \operatorname{Tr}\bigg[ 
	& \gamma_{\mu} \left ( \slashed{p}_{1}+m_{Q} \right) \gamma^{\mu} S^{t}_{a}(p_{1}-q_{1}) \gamma^{\nu} \left( \slashed{p}_{2} -m_{Q}\right) \gamma_{\nu} S^{t\dagger}_{a}(p_{1}-q_{1})  \nonumber \\
	& + \gamma_{\mu} \left ( \slashed{p}_{1}+m_{Q} \right) \gamma^{\mu} S^{t}_{b}(p_{1}-q_{1}) \gamma^{\nu} \left( \slashed{p}_{2} -m_{Q}\right) \gamma_{\nu} S^{t\dagger}_{b}(p_{1}-q_{1}) \nonumber \\
	& +2~ \text{Re}~\gamma_{\mu}\left ( \slashed{p}_{1}+m_{Q} \right) \gamma^{\mu} \left(i S^{t}_{b}(p_{1}-q_{1})  \right)  \gamma^{\nu} \left( \slashed{p}_{2} -m_{Q}\right) \gamma_{\nu} S^{t\dagger}_{a}(p_{1}-q_{1})   \bigg]
\end{align}
The cross-term between $S_{b}^{t}$ and $S_{a}^{t}$ is purely imaginary due to the explicit factor $i$ in $S_{b}^{t}$, and since the physical amplitude squared must be real, this term vanishes. The final form of the amplitude squared for the \(t-\)channel is given by 
\begin{align}
	\left | \mathcal{M}_{t} \right |^2 = g_{s}^{4} \mathcal{C}_{tt} \operatorname{Tr}\bigg[ 
	& \gamma_{\mu} \left ( \slashed{p}_{1}+m_{Q} \right) \gamma^{\mu} S^{t}_{a}(p_{1}-q_{1}) \gamma^{\nu} \left( \slashed{p}_{2} -m_{Q}\right) \gamma_{\nu} S^{t\dagger}_{a}(p_{1}-q_{1})  \nonumber \\
	& + \gamma_{\mu} \left ( \slashed{p}_{1}+m_{Q} \right) \gamma^{\mu} S^{t}_{b}(p_{1}-q_{1}) \gamma^{\nu} \left( \slashed{p}_{2} -m_{Q}\right) \gamma_{\nu} S^{t\dagger}_{b}(p_{1}-q_{1}) \bigg], \label{totalcrossampt}
\end{align}	
where 
\begin{align}
 S^{t}_{a}(p_{1}-q_{1}) = \frac{\left(\slashed{p}_{1}-\slashed{q}_{1}  \right)+m_{Q}}{t-m_{Q}^2}, \quad S^{t}_{b} (p_{1}-q_{1}) = e B\,\frac{ \left[ \left(\slashed{p}_{1}-\slashed{q}_{1}  \right)_\parallel+m_{Q} \right] \gamma^1\gamma^2}{(t-m_{Q}^2)^2}.
\end{align}
Next we consider the \(u-\)channel process, and we proceed similarly as before. The amplitude squared for the \(u-\)channel after omitting the imaginary part is 	
\begin{align}
	\left | \mathcal{M}_{u} \right |^2 = g_{s}^{4} \mathcal{C}_{uu} \operatorname{Tr}\bigg[ 
	& \gamma_{\nu} \left ( \slashed{p}_{1}+m_{Q} \right) \gamma^{\nu} S^{u}_{a}(q_{1}-p_{2}) \gamma^{\mu} \left( \slashed{p}_{2} -m_{Q}\right) \gamma_{\mu} S^{u\dagger}_{a}(q_{1}-p_{2})  \nonumber \\
	& + \gamma_{\nu} \left ( \slashed{p}_{1}+m_{Q} \right) \gamma^{\nu} S^{u}_{b}(q_{1}-p_{2}) \gamma^{\mu} \left( \slashed{p}_{2} -m_{Q}\right) \gamma_{\mu} S^{u\dagger}_{b}(q_{1}-p_{2}) \bigg], \label{totalcrossampu}
\end{align}	
where the colour factor is 	\(\mathcal{C}_{uu} = \sum_{\text{col}} \text{Tr} \left [t^{b}t^{a}t^{a}t^{b} \right ] = C_{A}C_{F}^{2} = \frac{16}{3}\) and 
\begin{align}
	S^{u}_{a}(q_{1}-p_{2}) = \frac{\left(\slashed{q}_{1}-\slashed{p}_{2}  \right)+m_{Q}}{u-m_{Q}^2}, \quad S^{u}_{b} (q_{1}-p_{2}) = e B\,\frac{ \left[ \left(\slashed{q}_{1}-\slashed{p}_{2}  \right)_\parallel+m_{Q} \right] \gamma^1\gamma^2}{(u-m_{Q}^2)^2}.
\end{align}	
The \(s-\)channel is unique to QCD, which contains a three-gluon vertex. 
The amplitude squared for the \(s-\)channel diagram is given by 
\begin{align}
	\left | \mathcal{M}_{s} \right |^2 = g_{s}^{4} \mathcal{C}_{ss} \text{Tr} \left[ \left( \slashed{p}_{1} +m_{Q}\right) \mathcal{K}^{\mu\nu}_{s} \left(\slashed{p}_{2} -m_{Q}\right) \mathcal{K}^{\dagger}_{s\mu\nu} \right] \label{totalcrossamps}
\end{align}
where the colour factor the \(s-\)channel is \(\mathcal{C}_{ss} = \sum_{\text{col}} f^{cab} f^{dab} \text{Tr} \left [t^{c}t^{d} \right ] = C_{A}^{2}C_{F} = 12 \) and 
\begin{align}
	\mathcal{K}^{\mu\nu}_{s} = \frac{1}{s} \left[ g^{\mu\nu} \left( \slashed{q}_{1}-\slashed{q}_{2}  \right ) -2\gamma^{\mu}q_{1}^{\nu}+2\gamma^{\nu}q_{2}^{\mu}\right].
\end{align}	
Next, we consider the cross-channel amplitudes and list the results:
\begin{align}
	\nonumber 2\text{Re}|\mathcal{M}_{t}\mathcal{M}^{\dagger}_{u}|   = 2 g_{s}^{4} \mathcal{C}_{tu} \text{Tr} \bigg[& \gamma_{\nu} \left(\slashed{p}_{1}+m_{Q}\right)\gamma^{\mu}S_{a}^{t}(p_{1}-q_{1}) \gamma^{\nu} \left(\slashed{p}_{2}-m_{Q} \right)\gamma_{\mu} S_{a}^{u \dagger}(q_{1}-p_{2}) \\ & + \gamma_{\nu} \left(\slashed{p}_{1}+m_{Q}\right)\gamma^{\mu}S_{b}^{t}(p_{1}-q_{1}) \gamma^{\nu} \left(\slashed{p}_{2}-m_{Q} \right)\gamma_{\mu} S_{b}^{u \dagger}(q_{1}-p_{2})  \bigg], \label{totalcrossamptu} \\
	2\text{Re}|\mathcal{M}_{t}\mathcal{M}^{\dagger}_{s}|   = 2 g_{s}^{4} \mathcal{C}_{ts} \text{Tr} \bigg[&  \left(\slashed{p}_{1}+m_{Q}\right)\gamma^{\mu}S_{a}^{t}(p_{1}-q_{1}) \gamma^{\nu} \left(\slashed{p}_{2}-m_{Q} \right) \mathcal{K}^{s \dagger}_{\mu\nu} \bigg], \label{totalcrossampts}\\
	2\text{Re}|\mathcal{M}_{u}\mathcal{M}^{\dagger}_{s}|   = 2 g_{s}^{4} \mathcal{C}_{us} \text{Tr} \bigg[&  \left(\slashed{p}_{1}+m_{Q}\right)\gamma^{\nu}S_{a}^{u}(p_{1}-q_{1}) \gamma^{\mu} \left(\slashed{p}_{2}-m_{Q} \right) \mathcal{K}^{s \dagger}_{\mu\nu} \bigg]\label{totalcrossampus},
\end{align}
for the cross-channels the colour factors are \(\mathcal{C}_{tu} = \sum_{\text{col}} \text{Tr} \left [t^{a}t^{b}t^{a}t^{b} \right ] =- \frac{C_{F}}{2} = -\frac{2}{3}\), \(\mathcal{C}_{ts} = \sum_{\text{col}} i f^{cab}\text{Tr} \left [t^{b}t^{a}t^{c} \right ]  = \frac{C^{2}_{A}C_{F}}{2}=6\) and \(\mathcal{C}_{us} = \sum_{\text{col}} i f^{abc}\text{Tr} \left [t^{c}t^{a}t^{b} \right ] = -\frac{C^{2}_{A}C_{F}}{2}=-6\). In $ts$ and $us$ cross-channels, the interference terms between $S_{b}^{t(u)}$ and $\mathcal{K}^{\dagger}_{\mu\nu}$ are purely imaginary $-$ owing to the explicit factor of $i$ in $S_{b}^{t(u)}$, and thus vanish upon taking the real part of the amplitude squared. The detailed expressions of invariant amplitudes after evaluating the Dirac traces are provided in Appendix \ref{Explicit expressions for the amplitude squared in a weak magnetic field}.
\subsection{Kinematic configuration in the presence of a magnetic field}
\label{Kinematic configuration in the presence of a magnetic field}
\begin{figure}[h]
 	\begin{minipage}{0.45\linewidth}
 		\includegraphics[scale=0.3]{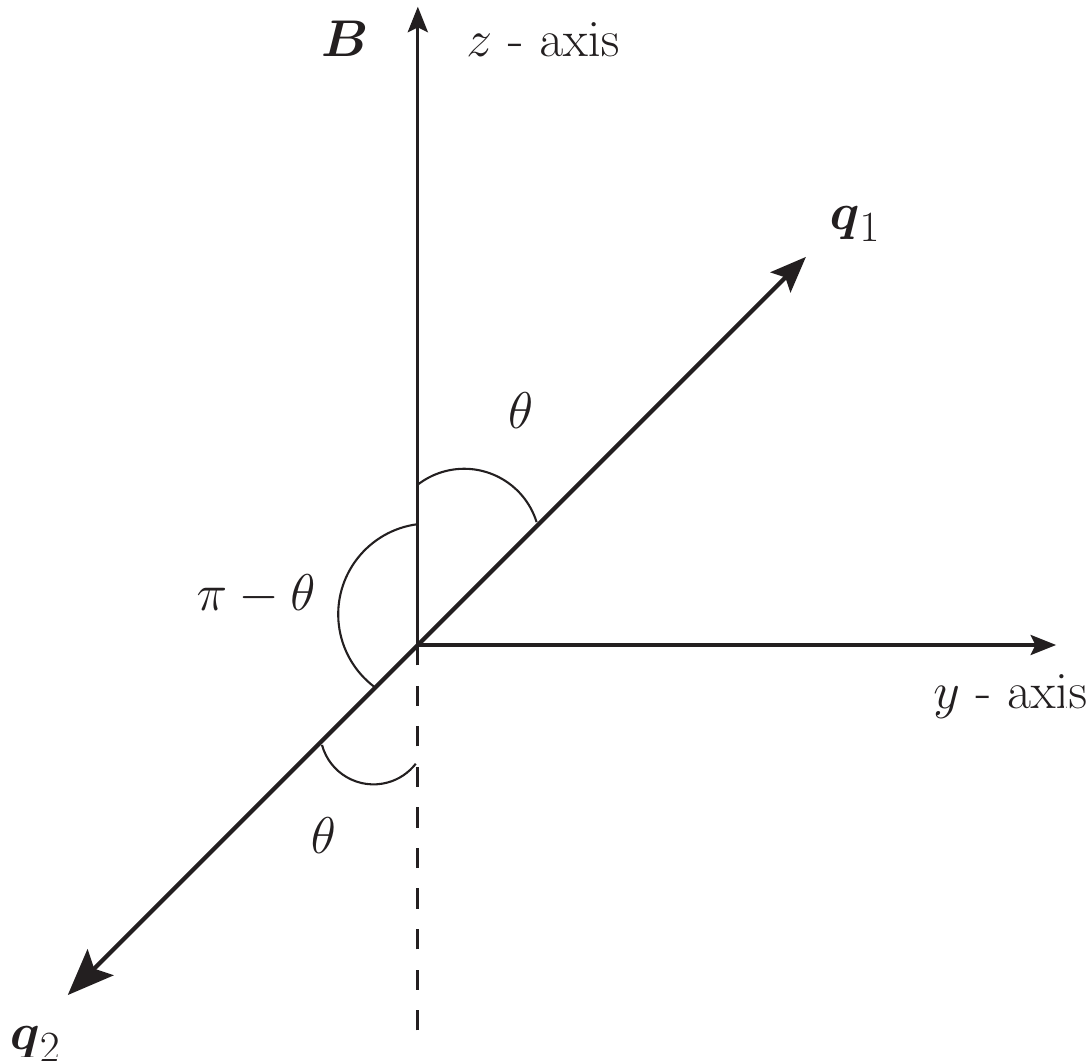}
 	\end{minipage}
 	\begin{minipage}{0.45\linewidth}
 		\includegraphics[scale=0.3]{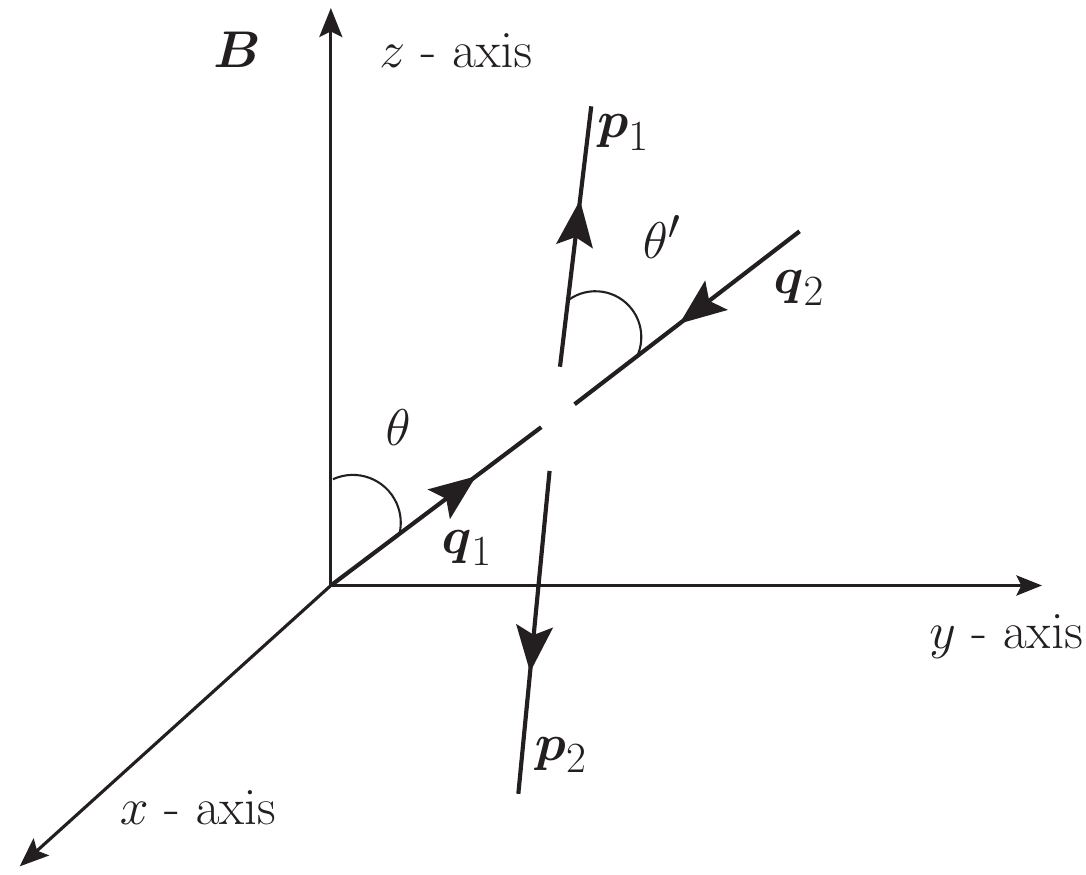}
 	\end{minipage}  
 	\caption{\textbf{Left panel}: Relative orientation between the momenta of the incoming gluons and the magnetic field. \textbf{Right panel}: Since there is no azimuthal dependence around the direction of the magnetic field ($z-$axis), we take the outgoing particle momenta in the $y-z$ plane, making a scattering angle $\theta^{\prime}$ with respect to the incoming gluon momentum vector.}
 	\label{graph}
 \end{figure}

In an experimental set-up, we take the beam axis to be along the $y-$axis. The magnetic field is along the $z-$axis, and the $x-y$	plane or the $z=0$ plane, defines the reaction plane. On the reaction plane, two heavy ions collide. In this calculation, we consider the interaction at the partonic level and want to calculate the probability amplitude of two gluons fusing together to produce a heavy quark-antiquark pair. Since the QCD interaction is taken into account, the partons may not have collinear motion with the parent nucleon or the heavy ion. In this case, the incoming gluons momenta can make an angle with respect to the direction of the magnetic field. Let us assume that this orientation angle is $\theta$, as shown in the left panel of Fig. \ref{graph}. Now the cross-section does not depend on the azimuthal angle around the z-axis. So, for simplicity we can take the three-momenta of the incoming gluons to be on the $y-z$ plane, with the vector making an angle $\theta$ with the $z-$axis. In the center of mass (CM) frame: 
\begin{align}
\nonumber	\boldsymbol{q}_{1} + \boldsymbol{q}_{2} =& \boldsymbol{p}_{1} + \boldsymbol{p}_{2}, \\
\nonumber	q_{1z} + q_{2z} =& p_{1z} + p_{2z}, \\
	\boldsymbol{q}_{1 \perp} + \boldsymbol{q}_{2 \perp} =& \boldsymbol{p}_{1 \perp} + \boldsymbol{p}_{2 \perp},
\end{align}
where $\boldsymbol{a}_{\perp} = \left(a^{1}, a^{2} \right)$ are the perpendicular components of the three vector $\boldsymbol{a}=\left(a^{1},a^{2},a^{3}\right)$.
 The CM energy is given by 
\begin{align}
	s = \left(q_{1}+q_{2}\right)^{2} = \left(q_{1}^{0} + q_{2}^{0} \right)^{2}.
\end{align}
So, without any loss of generality we can choose
\begin{align}
q_{1}^{0} = \frac{\sqrt{s}}{2}, \quad q_{2}^{0} = \frac{\sqrt{s}}{2}.
\end{align}
Similarly, we can also choose
\begin{align}
	p_{1}^{0} = \frac{\sqrt{s}}{2}, \quad p_{2}^{0} = \frac{\sqrt{s}}{2}.
\end{align}
On-shell conditions of the gluons and heavy quark pairs give us the following magnitudes of the three momentum vectors
\begin{align}
\nonumber	& \left|\boldsymbol{q}_{1}\right| = \frac{\sqrt{s}}{2}, \quad \left|\boldsymbol{q}_{2}\right| = \frac{\sqrt{s}}{2}, \\
	& \left|\boldsymbol{p}_{1}\right| = \sqrt{\frac{s}{4}-m_{Q}^2}, \quad \left|\boldsymbol{p}_{2}\right| = \sqrt{\frac{s}{4}-m_{Q}^2}.
\end{align}
The components of gluon three-momentum vectors along the $ z$-direction are given by 
\begin{align}
	q_{1z} = \frac{\sqrt{s}}{2}\cos{\theta}, \quad q_{2z} = -\frac{\sqrt{s}}{2}\cos{\theta}
\end{align}
From the on-shell conditions and the $z-$component of the momentum vectors of the gluons, we can find out the magnitudes of the parallel and perpendicular components of the gluon momenta four vectors and the results are
\begin{align}
\nonumber	&\left|\boldsymbol{q}_{1 \perp} \right| = \frac{\sqrt{s}}{2}\sin{\theta}, \quad \left|\boldsymbol{q}_{1 \parallel} \right| = \frac{\sqrt{s}}{2}\sin{\theta}, \\
	&\left|\boldsymbol{q}_{2 \perp} \right| = \frac{\sqrt{s}}{2}\sin{\theta}, \quad \left|\boldsymbol{q}_{2 \parallel} \right| = \frac{\sqrt{s}}{2}\sin{\theta}.
\end{align}
We use the convention written in Eq. \ref{convention} for various inner products. For the outgoing quark and anti-quark spinors, the effect of the magnetic field is neglected due to the scale hierarchy $\left(m_{Q}^{2} \gg eB \right)$. So in the final cross-section calculation, we integrate over the complete phase space of the outgoing particles.

Let us take $\theta^{\prime}$ as the scattering angle of the outgoing quark, as shown in the right panel of Fig. \ref{graph}. For simplicity we consider in plane scattering, i.e. we take the outgoing particle to be on the $y-z$ plane, making an angle $\theta^{\prime}$ with respect to the incoming gluon momentum vector. Because of the weak magnetic field approximation, we can use the usual four-momentum conservation and replace the outgoing anti-quark four-momentum with 
\begin{align}
	p_{2} = q_{1} + q_{2} -p_{1}.
\end{align}
So, we only need to determine the components of the three-momentum vector $\boldsymbol{p}_{1}$ of the outgoing quark. In order to determine the components of $\boldsymbol{p}_{1}$, we note that we can write a unit vector along $\hat{\boldsymbol{q}}_{1}$ and a unit vector $\hat{\boldsymbol{e}}_{1}$ perpendicular to that as 
\begin{align}
	\hat{\boldsymbol{q}}_{1} = \left(0, \sin{\theta}, \cos{\theta} \right), \quad \hat{\boldsymbol{e}}_{1} = \left(0, \cos{\theta}, -\sin{\theta} \right).
\end{align}
Together $\{\hat{\boldsymbol{q}}_{1}, \hat{\boldsymbol{e}}_{1}\}$ form an orthonormal system and a set of basis vectors. The components of $\boldsymbol{p}_{1}$ along $\hat{\boldsymbol{q}}_{1}$ and $\hat{\boldsymbol{e}}_{1}$ are given by
\begin{align}
	\boldsymbol{p}_{1} = \sqrt{\frac{s}{4}-m_{Q}^{2}} \left(\cos{\theta^{\prime}} \hat{\boldsymbol{q}}_{1} + \sin{\theta^{\prime}} \hat{\boldsymbol{e}}_{1}\right).
\end{align}
 So the component of $\boldsymbol{p}_{1}$ along the $z-$axis is 
 \begin{align}
\nonumber p_{1z} &= 	\boldsymbol{p}_{1} \cdot \hat{\boldsymbol{z}} \\ \nonumber &= \sqrt{\frac{s}{4}-m_{Q}^{2}} \left[\cos{\theta^{\prime}} \left(\hat{\boldsymbol{q}}_{1} \cdot \hat{\boldsymbol{z}} \right) + \sin{\theta^{\prime}} \left(\hat{\boldsymbol{e}}_{1} \cdot \hat{\boldsymbol{z}} \right)\right] \\  &= \sqrt{\frac{s}{4}-m_{Q}^{2}} \left(\cos{\theta^{\prime}} \cos{\theta} - \sin{\theta^{\prime}} \sin{\theta}\right) = \sqrt{\frac{s}{4}-m_{Q}^{2}} \cos{\alpha},
 \end{align}
where $\cos{\alpha} = \cos{\theta^{\prime}} \cos{\theta} - \sin{\theta^{\prime}} \sin{\theta}$. So, from the magnitude of $\boldsymbol{p}_{1}$ we can find the magnitude squared of the transverse component of $\boldsymbol{p}_{1}$ as
\begin{align}
	{p}_{1 \perp}^{2} = \left(\frac{s}{4} -m_{Q}^{2}\right) \left(1- \cos^{2} \alpha \right).
\end{align}
Similarly, from the on-shell condition of the heavy quark, we can find out 
\begin{align}
	{p}_{1 \parallel}^{2} = \frac{s}{4} - \left(\frac{s}{4} -m_{Q}^{2}\right) \cos^{2} \alpha.
\end{align}
The angle $\theta^{\prime}$ can also be replaced in terms of the Mandelstam variables in the following way. The Mandelstam variable $t$ can be defined as
\begin{align}
	m_{Q}^{2} - t = 2 q_{1} \cdot p_{1} = \frac{s}{2} \left(1 - \beta \cos{\theta^{\prime}} \right),
\end{align}
where we have defined $\beta = \sqrt{1- \frac{4m_{Q}^2}{s}}$. From this relation we can substitute the angle $\theta^{\prime}$ as 
\begin{align}
\nonumber	\cos^{2}{\theta^{\prime}} &= \left(\frac{s + 2 \left( t - m_{Q}^{2}\right)}{\beta s} \right)^{2}, \\ 
	\sin^{2}{\theta^{\prime}} &= 1 - \left(\frac{s + 2 \left( t - m_{Q}^{2} \right)}{\beta s} \right)^{2}.
\end{align}
With a non-zero magnetic field, the available four vectors are $\left(q_{1 \parallel}, q_{1 \perp}, q_{2 \parallel}, q_{2 \perp}, p_{1 \parallel}, p_{1 \perp}, p_{2 \parallel}, p_{2 \perp} \right)$. The last two four-vectors can be replaced by the conservation of four-momentum for the parallel and perpendicular components
\begin{align}
\nonumber	p_{2 \parallel} &= q_{1 \parallel} + q_{2 \parallel} - p_{1 \parallel}, \\
	p_{2 \perp} &= q_{1 \perp} + q_{2 \perp} - p_{1 \perp}.
\end{align}
So finally, we list all the possible Lorentz scalars that can be constructed out of these four-vectors
\begin{align}
\nonumber	q_{1 \parallel}^{2} &= \frac{s}{4} \sin^{2}{\theta}, \quad
	q_{1 \perp}^{2} = \frac{s}{4} \sin^{2}{\theta}, \quad
	q_{2 \parallel}^{2} = \frac{s}{4} \sin^{2}{\theta}, \quad
	q_{2 \perp}^{2} = \frac{s}{4} \sin^{2}{\theta}, \\
\nonumber	{p}_{1 \parallel}^{2} &= \frac{s}{4} - \left(\frac{s}{4} -m_{Q}^{2}\right) \cos^{2} \alpha, \quad
	{p}_{1 \perp}^{2} = \left(\frac{s}{4} -m_{Q}^{2}\right) \left(1- \cos^{2} \alpha \right), \\
\nonumber	q_{1 \parallel} \cdot q_{2 \parallel} &= \frac{s}{4}\left(1 + \cos^{2}{\theta} \right), \quad
	q_{1 \parallel} \cdot p_{1 \parallel} = \frac{s}{4} \left(1 - \beta \cos{\theta} \cos{\alpha} \right), \\
\nonumber	q_{1 \perp} \cdot q_{2 \perp} &= -\frac{s}{4}\sin^{2}{\theta}, \quad
	q_{1 \perp} \cdot p_{1 \perp} = \frac{s}{4} \beta \left(\cos{\theta^{\prime}} - \cos{\theta} \cos{\alpha} \right), \\ 
	q_{2 \parallel} \cdot p_{1 \parallel} &= \frac{s}{4} \left(1 + \beta \cos{\theta} \cos{\alpha} \right), \quad
	q_{2 \perp} \cdot p_{1 \perp} = \frac{s}{4}\beta \left( -\cos{\theta^{\prime}}+\cos{\theta} \cos{\alpha}\right). \label{lorentz scalars}
\end{align}
The Lorentz scalars listed above, together with the Mandelstam variables $s$, $t$ and $u$, form a complete set of kinematic invariants needed to express the cross-section. These will be substituted in the amplitude-squared expressions of the previous section to obtain the final total cross-section of for $g+g \rightarrow Q + \bar{Q}$ process in the presence of a weak magnetic field. This kinematic relations derived here make the anisotropic structure of the process explicit. In the absence of the magnetic field the cross-section depends only on the Mandelstam variables $s$, $t$ and $u$, which are rotaionally invariant. In the presence of the magnetic field, however, the parallel and perpendicular projections of the momenta enter independently, introducing explicit dependence on the orientation angle $\theta$ between the incoming gluons and the direction of the magnetic field. In the limit $\theta \rightarrow \pi/2$ the gluons propagate perpendicular to the magnetic field, while $\theta \rightarrow 0, \pi$ the corresponding propagation is parallel and anti-parallel to the magnetic field. These limiting cases will serve as useful checks on the final result.
\subsection{Analytical Expression of the Cross-section and Results}
\label{Cross-section for the process}
We now write the cross-section for the $g+g\to Q+\bar{Q}$ process in the CM frame. The total cross-section depends on the Mandelstam variable $s$, the magnetic field $B$ and the angle $\theta$ between the direction of the magnetic field and the incoming gluons momenta, as shown in Fig. \ref{graph}. The total cross-section $\sigma_{\text{tot}}(s,B,\theta)$ obtained from the invariant amplitude squared is given by
\begin{align}
	\sigma_{\text{tot}}(s,B,\theta) = \sigma_{0}(s) + \sigma_{\text{cor}}(s,B,\theta) = \frac{1}{16 \pi \lambda(s, 0, 0)}\int_{m_{Q}^{2}-\frac{s}{2}\left[1+\beta\right]}^{m_{Q}^{2}-\frac{s}{2}\left[1-\beta\right]} |\overline{\mathcal{M}}|^{2} ~ dt, \label{cross-section formula}
\end{align}
where $\lambda(x,y,z) = (x^2 + y^2 +z^2 -2xy - 2yz - 2zx)$ is the K\"all\'en function, $\beta = \sqrt{1- \frac{4m_{Q}^2}{s}}$ and the $|\mathcal{\overline{M}}|^{2}$ expression means that the amplitude squared is averaged over the initial gluon polarizations and colours, i.e. $|\mathcal{\overline{M}}|^{2} = \frac{1}{2 \cdot 2 \cdot 8 \cdot 8}|\mathcal{{M}}|^{2}$, with the amplitude squared is defined Eq. \ref{totalamplitudesquared} and detailed expression is given in Appendix from Eqs. \ref{totalamplitudesquaredappendix} - \ref{totalamplitudesquaredappendixlast}. We first consider the cross-section $\sigma_{0}(s)$, due to the first term of the propagator in Eq. \ref{propagator}, which is also the leading-order heavy quark production cross-section in vacuum. The expression for the leading-order cross-section is given by 
\begin{align}
	\sigma_{0}(s) = \frac{\pi \alpha_{s}^{2}}{3 s} \left[\left(1+\frac{4m_{Q}^{2}}{s}+\frac{m_{Q}^{4}}{s^{2}}\right)\log\left(\frac{1+\beta}{1-\beta}\right)-\left(\frac{31m_{Q}^{2}}{4s}+\frac{7}{4}\right)\beta\right], \label{leadingorderCS}
\end{align}
where $g_{s}^{2}=4\pi\alpha_{s}$, this expression agrees exactly with the Refs. \cite{Chen:2024lmp, Combridge:1978kx, Gluck:1977zm}. The expression is singular as $s\rightarrow0$, however, the physical threshold for heavy quark pair production is $\sqrt{s}=2m_{Q}$, which we take as the lower limit for numerical evaluation. Next, we consider the correction to the leading-order cross-section due to the second term of the propagator in Eq. \ref{propagator}, which is given by
\begin{align}
  \nonumber   \sigma_{\text{cor}}(s,B, \theta) =& \frac{\pi \alpha_{s}^2 \left(eB\right)^2}{144m_Q^4s}\bigg[\left(-3-\frac{14m_Q^2}{s}+\frac{53m_Q^4}{s^2} \right)\beta -\frac{6m_Q^6}{s^3} \log \left(\frac{1+\beta}{1- \beta} \right) + \frac{6m_Q^4}{s^2}\log\left( \frac{1+\beta}{1-\beta} \right) \\ \nonumber +& \cos 2\theta \left( \left( 20-\frac{88m_Q^2}{s}+\frac{80m_Q^4}{s^2} \right)\beta + \left( \frac{84m_Q^4}{s^2}+ \frac{72m_Q^6}{s^3} - \frac{72m_Q^{8}}{s^4} \right) \log \left( \frac{1+\beta}{1-\beta} \right) \right) \\ +& \cos{4 \theta} \left( \left( -1+\frac{54m_Q^2}{s}+\frac{439m_Q^4}{s^2}+\frac{12m_Q^6}{s^3} \right)\beta +\left( \frac{25m_Q^{8}}{s^4}-\frac{66m_Q^6}{s^3}-\frac{306m_Q^4}{s^2} \right)\log \left( \frac{1+\beta}{1-\beta} \right) \right) \bigg]. \label{correctionCS}
 \end{align}
 In the weak-field approximation, the magnetic field acts as a perturbation, and it is evident from Eq.~(\ref{correctionCS}) that the correction to the leading-order cross section scales as $(eB)^2$ rather than $(eB)$. This arises because the term $S_{1}(p)$ defined in Eq.~(\ref{correctionPropagator}), is linear in $(eB)$ and contributes to the amplitude squared only through $|S_{1}|^{2}$, since the interference terms between $S_{0}(p)$ and $S_{1}(p)$ are purely imaginary and vanish upon taking the real part of the amplitude squared, as discussed in subsection \ref{Amplitude squared of different channels}. We also write here the $\theta-$integrated cross-section for the correction term
\begin{align}
    \sigma_{\text{cor}}(s,B) = \frac{\pi \alpha_{s}^{2} \left(eB\right)^2}{144 m_Q^{4} s} \left[ \left( -3 - \frac{14m_Q^2}{s} + \frac{53 m_Q^4}{s^2} \right)\beta + \frac{6m_Q^4}{s^3} \left( s-m_Q^2 \right) \log\left( \frac{1+\beta}{1-\beta} \right) \right]. \label{correctionCSInt}
\end{align}
 The corrections to the leading-order cross-section for three different angles, $\theta =0$, $\theta = \frac{\pi}{2}$ and $\theta = \pi$ are given by
 \begin{align}
 	\sigma_{\text{cor}}(s,B,0) &=\sigma_{\text{cor}}(s,B,\pi)= \frac{\pi \alpha_{s}^{2} \left(eB\right)^{2}}{36m_Q^{4}s}\left[\left(4-\frac{12m_Q^{2}}{s}+\frac{143m_Q^{4}}{s^{2}}-\frac{6m_Q^{6}}{s^{3}}\right)\beta -6m_Q^{4} \left(\frac{2m_Q^{4}}{s^{4}}+\frac{9}{s^{2}}\right) \log\left(\frac{1+\beta}{1-\beta}\right) \right], \label{correctionCSzero}\\
 	\sigma_{\text{cor}}(s,B,\frac{\pi}{2}) &= \frac{\pi \alpha_{s}^{2} \left(eB\right)^{2}}{36m_Q^{4}s} \left[\left(-6+\frac{32m_Q^2}{s}+\frac{103m_Q^{4}}{s^{2}}+\frac{12m_Q^{6}}{s^{3}}\right)\beta +12m_Q^{4}\left(\frac{2m_Q^{4}}{s^{4}}-\frac{3m_Q^{2}}{s^{3}}-\frac{8}{s^{2}}\right)\log\left(\frac{1+\beta}{1-\beta}\right) \right],\label{correctionCS90}
 \end{align}
  From these expressions it is very clear that $\sigma_{\text{cor}}(s,B,0) = \sigma_{\text{cor}}(s,B,\pi)$. This equality follows directly from Eq.~(\ref{correctionCS}), where the $\theta$-dependence enters only through $\cos{2\theta}$ and $\cos{4\theta}$, both of which takes the same value at $\theta=0$ and $\theta=\pi$. 
\section{Numerical analysis}
\label{Numerical analysis}
\begin{figure}[h]
	\includegraphics[scale=0.6]{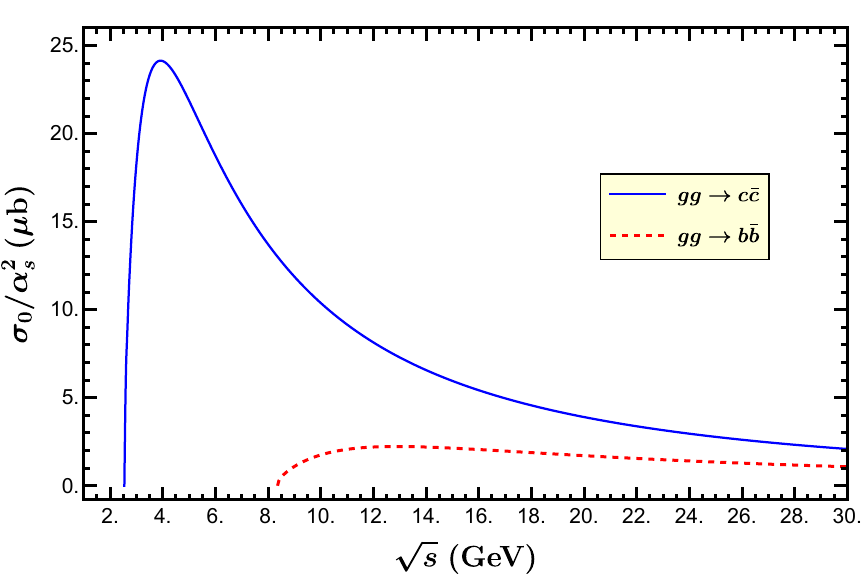}
	\caption{The leading-order scaled cross-section as a function of $\sqrt{s}$ for the heavy quark production process \(g (q_{1})+g(q_{2}) \to Q(p_{1})+\bar{Q}(p_{2})\) for the charm ($m_{Q}= 1.27 \text{GeV}$) and bottom ($m_{Q}= 4.18 \text{GeV}$) quarks.}
	\label{plotTreeLevel}
\end{figure}

\begin{figure}[h]
	\begin{minipage}{0.45\linewidth}
		\includegraphics[scale=0.55]{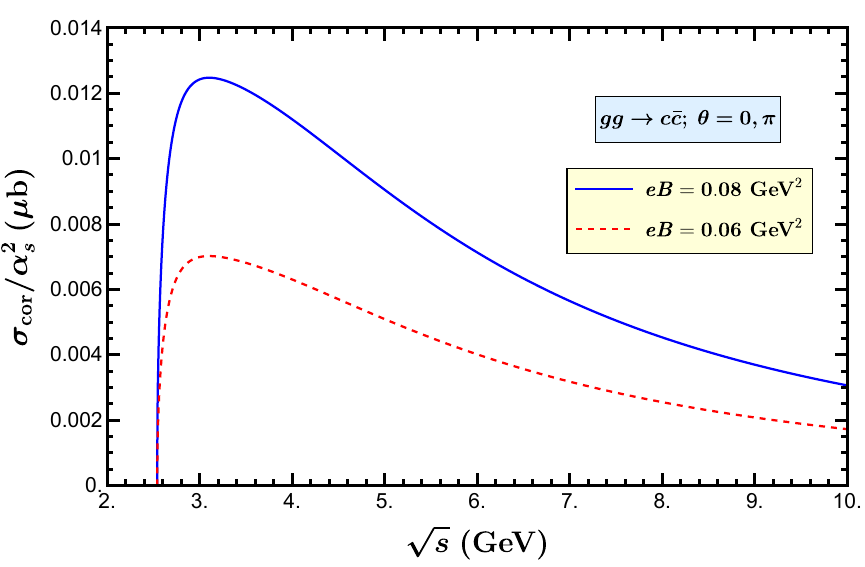}
	\end{minipage}
	\begin{minipage}{0.45\linewidth}
		\includegraphics[scale=0.55]{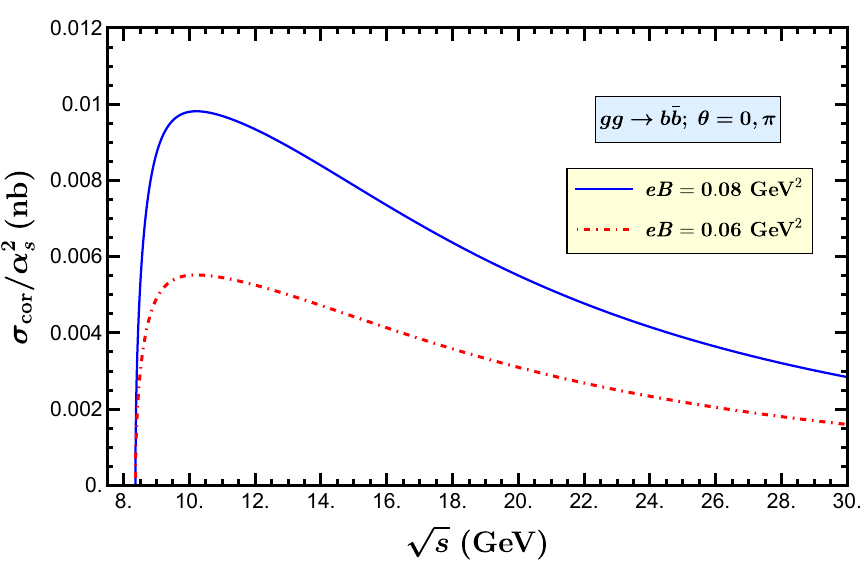}
	\end{minipage}  
	\caption{\textbf{Left panel}: The scaled correction to the leading order cross-section as a function of $\sqrt{s}$ for $\theta = 0,\pi$ and fixed magnetic field $eB$ for charm quark pair. \textbf{Right panel}: The scaled correction to the leading order cross-section as a function of $\sqrt{s}$ for $\theta = 0,\pi$ and fixed magnetic field $eB$ for bottom quark pair.}
	\label{corrections1}
\end{figure}

\begin{figure}[h]
	\begin{minipage}{0.45\linewidth}
		\includegraphics[scale=0.55]{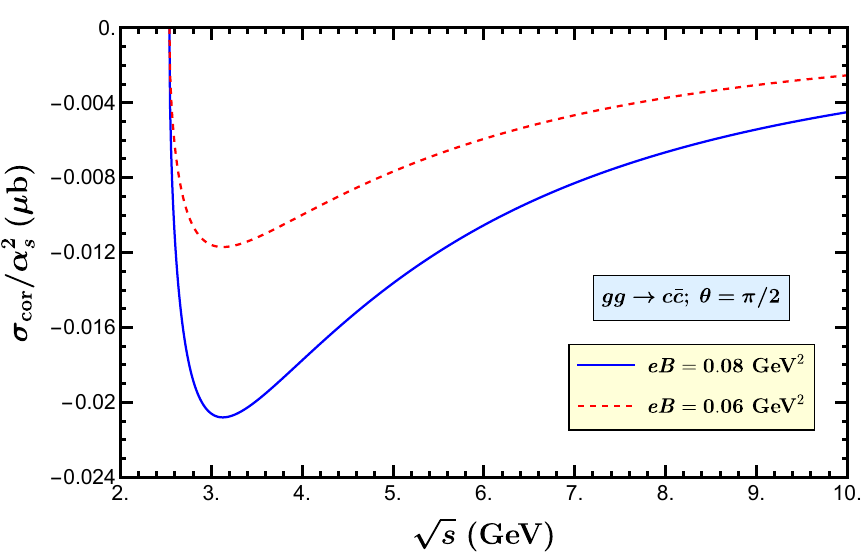}
	\end{minipage}
	\begin{minipage}{0.45\linewidth}
		\includegraphics[scale=0.55]{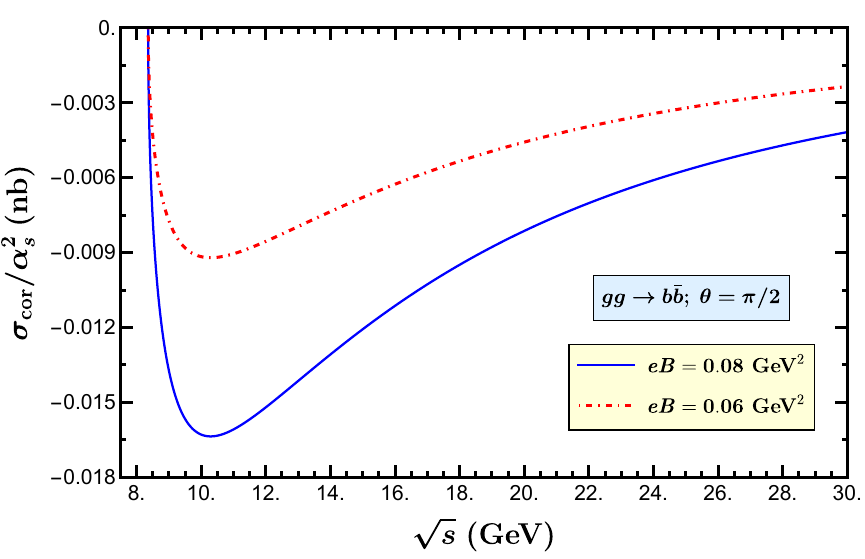}
	\end{minipage}  
	\caption{\textbf{Left panel}: The scaled correction to the leading order cross-section as a function of $\sqrt{s}$ for $\theta = \pi/2$ and fixed magnetic field $eB$ for charm quark pair. \textbf{Right panel}: The scaled correction to the leading order cross-section as a function of $\sqrt{s}$ for $\theta = \pi/2$ and fixed magnetic field $eB$ for bottom quark pair.}
	\label{corrections2}
\end{figure}

Plots of this leading-order scaled cross-section $\sigma_{0}/\alpha_{s}^{2}$ (to avoid coupling constant dependence) as a function of $\sqrt{s}$ for the charm ($m_{Q}= 1.27 ~\text{GeV}$) and bottom ($m_{Q}= 4.18 ~\text{GeV}$) quarks is shown in Fig. \ref{plotTreeLevel}. The cross-section for charm quark pair production is significantly higher than the bottom quark pair production because of the lower charm quark mass, thus requires less center-of-mass energy for a gluon pair to create a $c\overline{c}$ pair.
Plots of the scaled correction to the leading-order cross-section for angles $\theta = 0, \pi$ and $\theta=\pi/2$ as a function of $\sqrt{s}$ and two different values of the magnetic field $eB$ is shown in Fig. \ref{corrections1} and Fig. \ref{corrections2} both charm and bottom quark pairs. Although the typical magnetic field is $eB \sim m_{\pi}^{2} \approx 0.02 ~\text{GeV}^{2}$, we choose somewhat larger values $eB=0.08~ \text{GeV}^{2}$ and $0.06~\text{GeV}^{2}$ for illustrative purposes while remaining well within the weak field regime $eB \ll m_{Q}^{2}$.
We can see from the plots as well as expressions in Eqs. \ref{leadingorderCS}-\ref{correctionCS90} that at $\sqrt{s}=2m_{Q}$ the leading-order cross-section as well as the corrections become zero due to the presence of the term $\beta = \sqrt{1-\frac{4m_{Q}^{2}}{s}}$ in the expression. The corrections at angles $\theta=0$ and $\theta=\pi$ are found to be the same and positive; however, the correction term for $\theta=\pi/2$ is found to be negative. We note that the magnitude of the corrections are way less than the magnitude of the leading-order cross-section. For the bottom quark pair, the corresponding correction is of the order of nanobarns. So the total cross-section behaviour does not change much in the presence of a weak magnetic field; that's why we decided to show the correction terms to the leading-order cross-section separately. As we increase the magnetic field, the magnitude of the corrections increases, but it is still way less than the leading-order cross-section magnitude. We next consider the plots of the $\theta$-integrated correction to the leading-order cross-section shown in Eq. \ref{correctionCSInt}.
\begin{figure}[h]
	\begin{minipage}{0.45\linewidth}
		\includegraphics[scale=0.55]{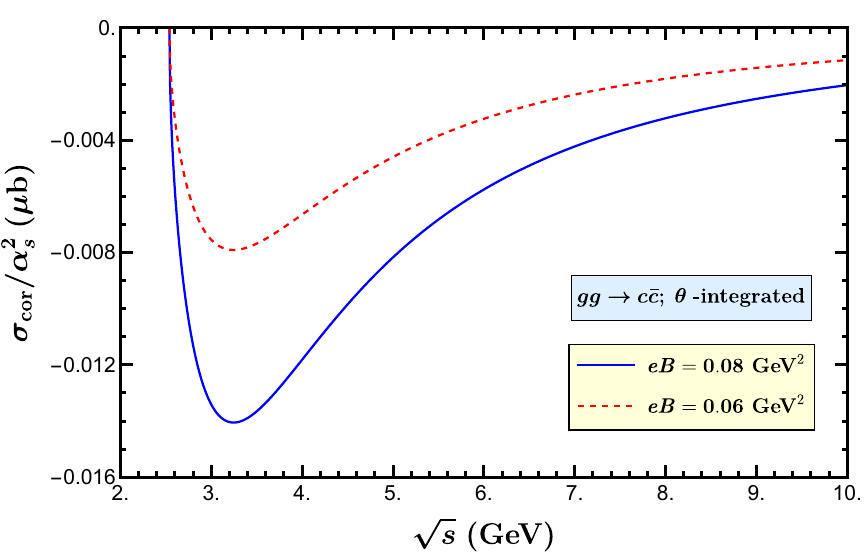}
	\end{minipage}
	\begin{minipage}{0.45\linewidth}
		\includegraphics[scale=0.55]{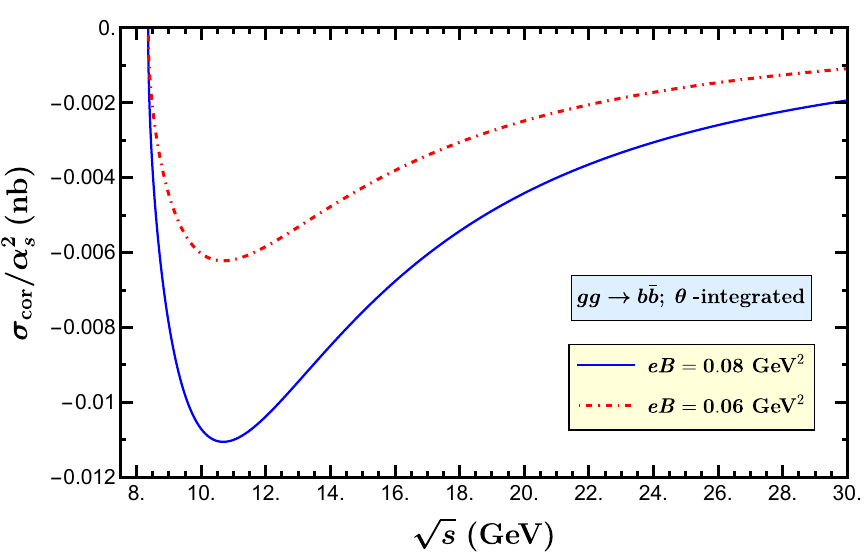}
	\end{minipage}  
	\caption{\textbf{Left panel}: The scaled $\theta$-integrated correction to the leading order cross-section as a function of $\sqrt{s}$ and fixed magnetic field $eB$ for charm quark pair. \textbf{Right panel}: The scaled $\theta$-integrated correction to the leading order cross-section as a function of $\sqrt{s}$ and fixed magnetic field $eB$ for bottom quark pair.}
	\label{correctionsInt}
\end{figure}
The scaled $\theta$-integrated correction to the leading-order cross-section is shown in Fig. \ref{correctionsInt} for two different magnetic fields. This term is found to be negative, and the magnitude is very small compared to the leading-order cross-section magnitude for both the charm and the bottom quark pairs. As we increase the magnetic field, the negative magnitude increases as expected; however, the magnitude is still very small compared to the leading order. It is evident from the numerical evaluations that the magnitude of the cross-section for the gluon fusion process does not change much in the presence of a weak background magnetic field. 

\begin{figure}[h]
	\begin{minipage}{0.45\linewidth}
		\includegraphics[scale=0.55]{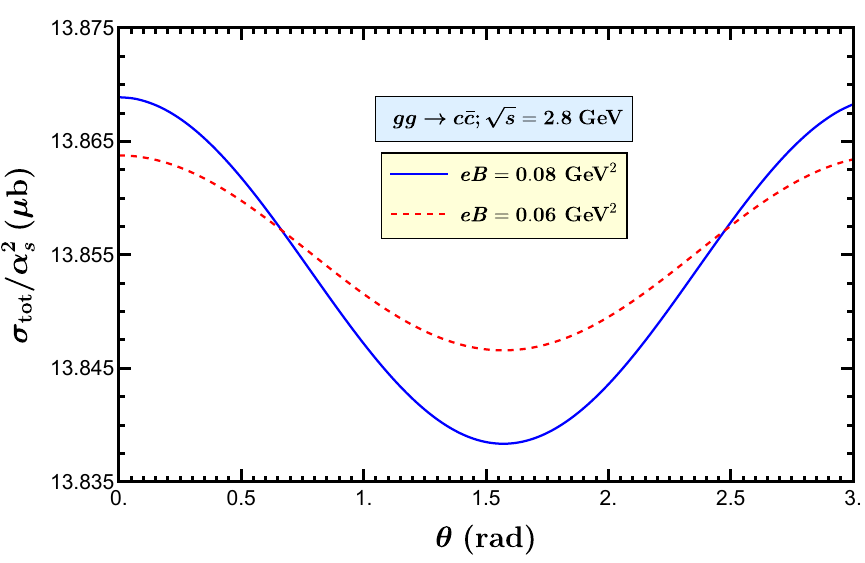}
	\end{minipage}
	\begin{minipage}{0.45\linewidth}
		\includegraphics[scale=0.55]{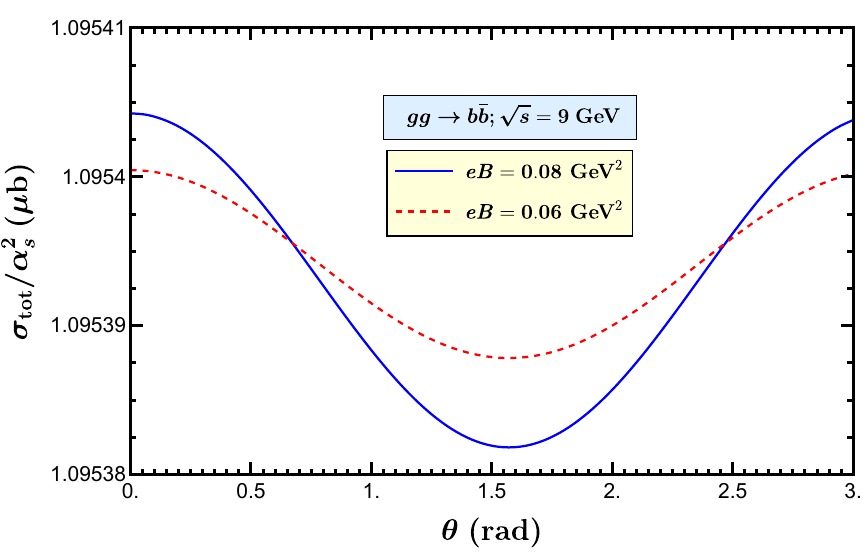}
	\end{minipage}  
	\caption{\textbf{Left panel}: The scaled total cross-section as a function of $\theta$ at fixed invariant mass $\sqrt{s}$ for charm quark pair. \textbf{Right panel}: The scaled total cross-section as a function of $\theta$ at fixed invariant mass $\sqrt{s}$ for bottom quark pair.}
	\label{AngDepWithB}
\end{figure}

The $\theta$-dependence of the total scaled cross-section at fixed invariant mass $\sqrt{s}$ is shown in Fig. \ref{AngDepWithB} for two values of magnetic field.
\begin{figure}[h]
	\begin{minipage}{0.45\linewidth}
		\includegraphics[scale=0.55]{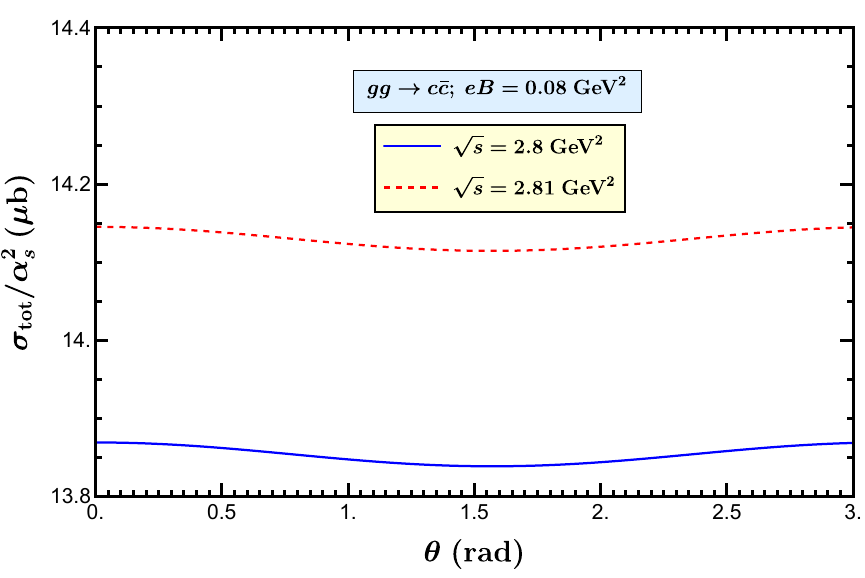}
	\end{minipage}
	\begin{minipage}{0.45\linewidth}
		\includegraphics[scale=0.55]{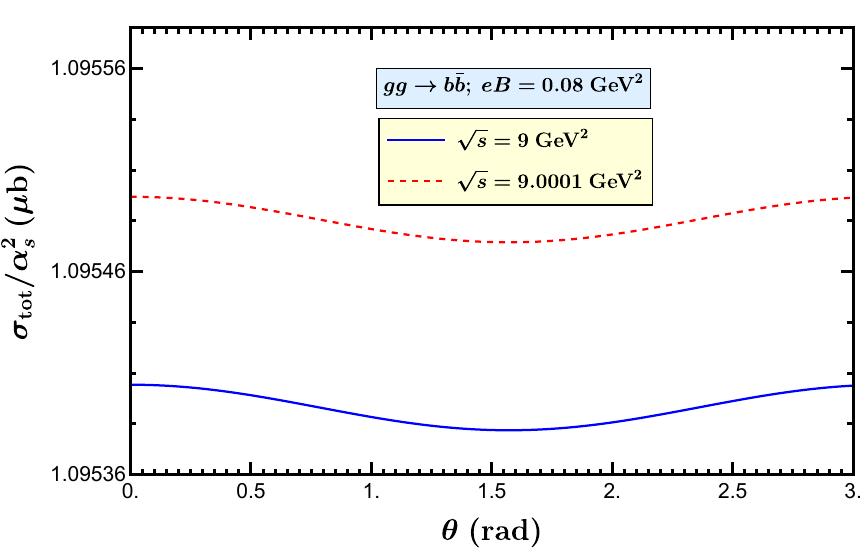}
	\end{minipage}  
	\caption{\textbf{Left panel}: The scaled total cross-section as a function of $\theta$ at fixed magnetic field $eB$ for charm quark pair. \textbf{Right panel}: The scaled total cross-section as a function of $\theta$ at fixed magnetic field $eB$ for bottom quark pair.}
	\label{AngDepWithS}
\end{figure}
\begin{figure}[h]
	\begin{minipage}{0.45\linewidth}
		\includegraphics[scale=0.55]{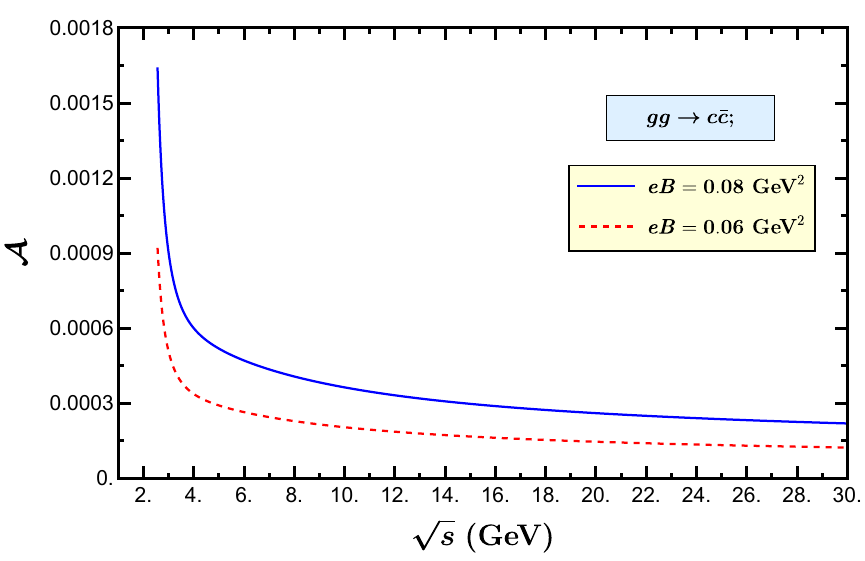}
	\end{minipage}
	\begin{minipage}{0.45\linewidth}
		\includegraphics[scale=0.55]{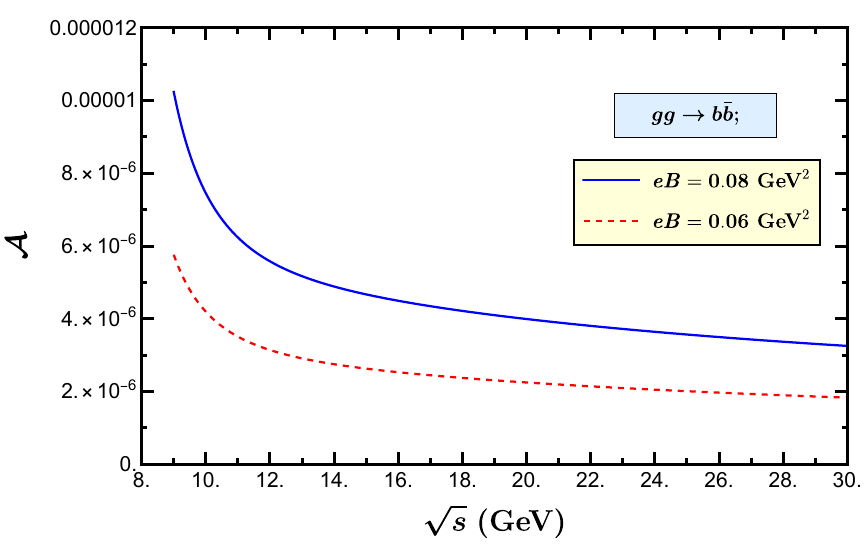}
	\end{minipage}  
	\caption{\textbf{Left panel}: The asymmetry function as a function of $\sqrt{s}$ for charm quark pair production. \textbf{Right panel}: The asymmetry function as a function of $\sqrt{s}$ for bottom quark pair production.}
	\label{plotAsymmFunc}
\end{figure}
We have taken the invariant mass little above the threshold for both charm $(\sqrt{s}=2.8 ~ \text{GeV})$ and bottom $(\sqrt{s}=9 ~ \text{GeV})$ quark pair production. The cross-section exhibits a $\pi$-periodic pattern, consistent with the $\cos{2\theta}$ and $\cos{4\theta}$ structure of the $\sigma_{\text{cor}}(s,\theta)$ in Eq.~(\ref{correctionCS}), and is symmetric about $\theta = \pi/2$. The maximum occurs at $\theta=0$ and $\theta=\pi$, corresponding to gluon momenta aligned with the magnetic field, while the minimum occurs at $\theta=\pi/2$, where gluons propagate perpendicular to the magnetic field. As expected from the $(eB)^{2}$ scaling of the correction, the amplitude of the angular variation grows with increasing magnetic field strength. The relative anisotropy, $\left[\sigma_{\text{tot}}(s,B,0)-\sigma_{\text{tot}}(s,B,\pi/2)\right]$, provides a gauge-invariant measure of the magnetic field effect and could in principle serve as an experimental signature of the initial-stage magnetic field in heavy-ion collisions. The $\theta$-dependence of the total scaled cross-section at a fixed magnetic field $(eB=0.08~\text{GeV}^2)$ for charm and bottom quarks is shown in Fig. \ref{AngDepWithS} for two values of invariant mass $\sqrt{s}$, both chosen near the production threshold. If we choose an invariant mass away from the production threshold, the angular variation practically vanishes. Thus, for illustrative purposes, we have taken the values of the invariant masses for both the charm and bottom quark pair near their production threshold. The angular variation patten with maximum at $\theta=0, \pi$ and minimum at $\theta=\pi/2$, is qualitatively identical for both values of $\sqrt{s}$. It is evident from Eqs. \ref{leadingorderCS} and \ref{correctionCS} that the anisotropy $\left[\sigma_{\text{tot}}(s,B,0)-\sigma_{\text{tot}}(s,B,\pi/2)\right]$, which is solely governed by the coefficient of $\cos{2\theta}$ remains approximately constant across the two chosen values of $\sqrt{s}$, i.e. it varies slowly with the invariant mass. Thus, Fig.~\ref{AngDepWithS} illustrates that the anisotropic structure of the cross section is a robust feature of the magnetic-field correction, preserved across a range of invariant masses near threshold.

To quantify the angular anisotropy induced by the magnetic field, we define the asymmetry function
\begin{align}
    \mathcal{A}(s,B)=\frac{\left[\sigma_{\text{tot}}(s,B,0)-\sigma_{\text{tot}}(s,B,\pi/2)\right]}{\left[\sigma_{\text{tot}}(s,B,0)+\sigma_{\text{tot}}(s,B,\pi/2)\right]},
\end{align}
which measures the relative difference between the cross sections at the two extremal orientations, $\theta=0$ and $\theta=\pi/2$. This will provide us with a range of invariant masses over which the anisotropic structure of the cross-section is preserved. The plot of this as a function of $\sqrt{s}$ is shown in Fig. \ref{plotAsymmFunc} for both charm and bottom quark pair production.
The asymmetry peaks nears the production threshold $\sqrt{s} \approx 2m_{Q}$, and decreases with increasing $\sqrt{s}$, for both values of the magnetic field. The asymmetry is significantly larger for the charm than for the bottom quark pair production. This is expected behaviour,since the asymmetry $\left[\sigma_{\text{tot}}(s,B,0)-\sigma_{\text{tot}}(s,B,\pi/2)\right]$ scales as $(eB)^{2}/m_{Q}^{4}$, see Eqs. \ref{leadingorderCS} and \ref{correctionCS}. Thus heavier quarks are considerably less sensitive to the magnetic field, as expected from the hierarchy $eB \ll m_{Q}^{2}$ that underlies the weak-field approximation. As the magnetic field increases, the asymmetry increases, which is a direct consequence of the fact that $\sigma_{\text{cor}} \propto (eB)^{2}$ for a fixed quark flavour. These results suggest that experimental searches for magnetic field-induced anisotropy in heavy quark production should focus on the near-threshold kinematic regime, where the asymmetry is maximized. The charm quark channel is considerably more sensitive than the bottom channel due to the $m_{Q}^{-4}$ suppression of the effect.

We note that all the numerical results presented above are obtained in the absence of a medium. In realistic heavy-ion collisions, however, the produced heavy quarks interact with the QGP medium, which modifies both the production dynamics and the subsequent evolution of the heavy quarks. The presence of the medium affects the magnetic field itself through the finite electrical conductivity of the QGP, which slows the decay of the field and prolongs its lifetime \cite{Tuchin:2013ie, Tuchin:2013apa, Tuchin:2014hza, Mustafa:2025uad}, and also modifies the fermion propagator through thermal corrections. A medium-modified analysis would require incorporating the finite-temperature fermion propagator in the Schwinger proper-time formalism \cite{Bandyopadhyay:2017cle, Das:2017vfh}, which introduces additional dependence on the temperature and the chemical potential. Such an extension is left for future work.
\section{Summary and Conclusions}
\label{sec:conclusion}
In this work, we have calculated the cross section for heavy quark pair production via the gluon fusion process $g+g \rightarrow Q+\bar{Q}$ in the presence of a weak external magnetic field, motivated by the intense magnetic fields generated in the early stages of non-central heavy-ion collisions at RHIC and LHC. The calculation is carried out at leading order in perturbative QCD, with the effect of the magnetic field incorporated through a perturbative expansion of the fermion propagator in Schwinger's proper-time formalism.

We neglect medium effects for simplicity. Since the vacuum magnetic field decays rapidly, we treat it as weak and constant in our gluon fusion calculation. The produced heavy quarks are also much heavier than the magnetic field scale ($m^2_Q \gg eB$, where $eB\sim m_\pi^2\sim 10^{18}~\text{G}$) at RHIC and LHC, further justifying the weak-field approximation. Under this approximation, the full Schwinger propagator in a constant magnetic field is expanded perturbatively, yielding the free Dirac propagator $S_{0}(p)$ as the leading term and a correction $S_{1}(p) \propto eB$ as the leading magnetic contribution. The effect of the magnetic field on the external heavy quark and antiquark spinors is neglected, since $\sqrt{eB} \ll m_{Q}$.

The total cross section decomposes naturally as $\sigma_{\text{tot}}(s,B,\theta) = \sigma_{0}(s) + \sigma_{\text{cor}}(s,B,\theta)$ where $\sigma_{0}(s)$ is the standard leading-order vacuum cross section $\sigma_{\text{cor}}(s,B,\theta)$ is the magnetic correction. The correction to the leading-order cross-section scales as $(eB)^{2}$ rather than $(eB)$. This arises because the leading magnetic term $S_{1}(p)$ in the propagator is linear in $eB$ and contribute to the amplitude squared through $|S_{1}|^{2} \propto (eB)^{2}$.Also, the interference terms between $S_{0}(p)$ and $S_{1}(p)$ are purely imaginary and vanish upon taking the real part of the amplitude squared.

The presence of the magnetic field breaks the rotational symmetry of the process
, rendering the cross section anisotropic. The magnetic correction depends on the orientation angle $\theta$ between the incoming gluon momenta and the magnetic field direction through $\cos{2\theta}$ and $\cos{4 \theta}$ terms. Consequently, the cross section is maximized when the incoming gluons propagate along the magnetic field direction $(\theta = 0, \pi)$ and minimized when they propagate perpendicular to it $(\theta=\pi/2)$. The equality $\sigma_{\text{cor}}(s,B,0)=\sigma_{\text{cor}}(s,B,\pi)$ follows directly from the periodicity of $\cos{2\theta}$ and $\cos{4\theta}$. The correction at $\theta=\pi/2$ is found to be negative, meaning that the magnetic field suppresses the cross section in this configuration relative to the vacuum result, while it enhances it at $\theta=0$ and $\theta=\pi$. The $\theta$-integrated correction $\sigma_{\text{cor}}(s,B)$ is found to be negative for all values of $\sqrt{s}$, indicating a net suppression of the total angle-averaged cross section in the presence of the magnetic field. The magnitude of the correction to the total cross-section induced by the weak magnetic field is found to be small compared to the leading-order cross-section.

To quantify the anisotropy, we have defined an asymmetry function, which measures the relative difference between the cross sections at the two extremal orientations. This function peaks at the production threshold $\sqrt{s}= 2m_{Q}$ and decreases with increasing invariant mass. The asymmetry scales as $(eB)^{2}/m_{Q}^{4}$, which makes it more sensitive to the charm quark channel than the bottom quark. These results constitute a signature of the initial-stage magnetic field that can, in principle, be imprinted on heavy-flavour observables measured in the final state.

\section*{ACKNOWLEDGMENTS}
U.K.D. acknowledges support from the Anusandhan National Research Foundation (ANRF), Government of India, under Grant Reference No. CRG/2023/003769. S.G. is funded by the Department of Higher Education, Government of West Bengal, India.

\appendix

\section{Explicit expressions for the amplitude squared in a weak magnetic field}
\label{Explicit expressions for the amplitude squared in a weak magnetic field}
In Sec.~\ref{Amplitude squared of different channels}, we listed the amplitude squared expressions for the different channels. The 
magnetic field introduces a preferred spatial direction — namely, the direction of $\boldsymbol{B}$ along the $z$-axis — thereby 
breaking the rotational symmetry of the process. As a consequence, the parallel (along $\boldsymbol{B}$) and perpendicular (transverse 
to $\boldsymbol{B}$) components of the momenta enter the expressions through the magnetic correction to the propagator (see Eq.~\ref{propagator}) and through the kinematic Lorentz scalars (see Eq.~\ref{lorentz scalars}). In evaluating the traces for 
the different amplitude-squared channels, these parallel and perpendicular components appear via the Lorentz scalars, which can be expressed in terms of the Mandelstam variables $(s, t, u)$ and the orientation angle $\theta$. We list the final expressions for the amplitude squared of the different channels below, where we have used $C_F = \frac{4}{3}$ and $C_A = 3$. The expression for $|\mathcal{M}_{s}|^{2}$, $|\mathcal{M}_{t}|^{2}$ and $|\mathcal{M}_{u}|^{2}$ are given by
\begin{align}
 |\mathcal{M}_{s}|^{2} &=    \pi ^2 \alpha ^2 C_A^2 {C_F}\frac{ \left(m^2-t\right) \left(-m^2+s+t\right)}{s^2},\\
 \nonumber |\mathcal{M}_{t}|^{2} &= \pi ^2 \alpha ^2C_{A}C_{F}^{2}\bigg[\frac{  \left(-3 m^4+m^2 s-t (s+t)\right)}{2 \left(t-m^2\right)^2}+\frac{(eB)^2}{16 s^2 \left(t-m^2\right)^4}\bigg\{s^2 \left(-14 m^4-16 m^2 t+t (3 s+2 t)\right)\\ \nonumber & +4 s \cos 2 \theta  \left(8 m^6-m^4 (s+20 t)+2 m^2 \left(s^2+4 s t+8 t^2\right)-t (s+t) (s+4 t)\right)+\cos 4 \theta  \big(32 m^8-16 m^6 (s+8 t)\\ \nonumber &+2 m^4 \left(s^2+40 s t+96 t^2\right)-16 m^2 t \left(s^2+7 s t+8 t^2\right)+t (s+2 t) \left(s^2+16 s t+16 t^2\right)\big)\\ \nonumber & +4 \sin 2 \theta  \sqrt{2 m^2 t-m^4-t (s+t)} \big(\cos 2 \theta  \left(-4 m^2+s+4 t\right) \left(-4 m^4+m^2 (s+8 t)-t (3 s+4 t)\right)\\&+s \left(8 m^4-m^2 (s+12 t)+t (3 s+4 t)\right)\big)\bigg\}\bigg],\\
    \nonumber |\mathcal{M}_{u}|^{2} &=\pi ^2 \alpha ^2C_{A}C_{F}^{2} \bigg[\frac{ \left(-7 m^4+m^2 (3 s+4 t)-t (s+t)\right)}{2 \left(-m^2+s+t\right)^2}+\frac{ (eB)^2}{16 s^2 \left(-m^2+s+t\right)^4}\bigg\{s^2 \big(-38 m^4+2 m^2 (7 s+4 t)\\ \nonumber &-(s-2 t) (s+t)\big)+4 s \cos 2 \theta  \left(-m^4 (s-4 t)+4 m^2 (s+t) (s-2 t)+t (s+t) (3 s+4 t)\right)+\cos 4 \theta  \big(32 m^8\\ \nonumber &-16 m^6 (3 s+8 t)+2 m^4 \left(13 s^2+88 s t+96 t^2\right)-2 m^2 \left(3 s^3+44 s^2 t+104 s t^2+64 t^3\right)\\ \nonumber &+(s+t) (s+2 t) \left(s^2+16 s t+16 t^2\right)\big)+4 \sqrt{-m^4+2 m^2 t-t (s+t)}\big(-s \left(m^2 (s-4 t)+(s+t) (s+4 t)\right)\\  &+\cos 2 \theta  +\left(4 m^2-3 s-4 t\right) \left(4 m^4-m^2 (3 s+8 t)+(s+t) (s+4 t)\right)\big)\sin 2\theta\bigg\}\bigg].
\end{align}
Whereas the final expressions for the cross terms come out to be
\begin{align}
\nonumber 2\text{Re}|\mathcal{M}_{t}\mathcal{M}^{\dagger}_{u}| &=\pi ^2 \alpha ^2C_{F} \bigg[\frac{ m^2 \left(s-4 m^2\right)}{2 \left(m^2-t\right) \left(m^2-s-t\right)}-\frac{(eB)^2}{8 s^2 \left(m^2-t\right)^2 \left(-m^2+s+t\right)^2}\bigg\{s^2 \left(2 m^4-m^2 (s+4 t)+2 t (s+t)\right)\\ \nonumber& + 4 s \cos 2 \theta  \left(6 m^6-4 m^4 (s+3 t)+m^2 \left(s^2+8 s t+6 t^2\right)-s t (s+t)\right) +\cos 4 \theta  \big(-32 m^8+32 m^6 (s+4 t)\\ \nonumber & -2 m^4 \left(5 s^2+64 s t+96 t^2\right)+m^2 \left(s^3+44 s^2 t+160 s t^2+128 t^3\right)-2 t (s+t) (s+4 t) (3 s+4 t)\big)\\ \nonumber & -2 \sin 2 \theta  \left(2 m^2-s-2 t\right) \sqrt{-m^4+2 m^2 t-t (s+t)} \big(\cos 2 \theta  \left(16 t \left(s-2 m^2\right)+\left(s-4 m^2\right)^2+16 t^2\right)\\ \nonumber & +s \left(s-6 m^2\right)\big)\bigg\} \bigg]  \\ 
2\text{Re}|\mathcal{M}_{t}\mathcal{M}^{\dagger}_{s}| &= \pi ^2 \alpha ^2 C_A^2 {C_F} \frac{ \left(t (s+t)-m^4\right)}{2 s \left(m^2-t\right)} \\
 2\text{Re}|\mathcal{M}_{u}\mathcal{M}^{\dagger}_{s}| &=\pi ^2 \alpha ^2 C_A^2 {C_F} \frac{ \left(3 m^4-2 m^2 (s+2 t)+t (s+t)\right)}{2 s \left(-m^2+s+t\right)}
\end{align}
It is evident from these expressions that the magnetic correction enters only through the $t$- and $u$-channel amplitudes and their 
mutual interference, since all other contributions are purely imaginary and vanish upon taking the real part of the amplitude squared, as discussed in Sec.~\ref{Amplitude squared of different channels}. To compute the total cross section, we group the total amplitude squared as follows:
\begin{align}
    |\mathcal{M}|^2 = |\mathcal{M}|_{(0)}^2 +|\mathcal{M}|_{(1)}^2++|\mathcal{M}|_{(2)}^2++|\mathcal{M}|_{(3)}^2. \label{totalamplitudesquaredappendix}
\end{align}
The first term $|\mathcal{M}|_{(0)}^2$ is the amplitude squared arising from the vacuum part $S_0(p)$ of the propagator in Eq.~\ref{LOpropagator}, and is given by
\begin{align}
  \nonumber  |\mathcal{M}|_{(0)}^2 &=\frac{2 \pi ^2 \alpha^{2}}{3 s^2 \left(m^2-t\right)^2 \left(-m^2+s+t\right)^2}  \left(-9 m^4+9 m^2 (s+2 t)-4 s^2-9 s t-9 t^2\right) \\ & \times \left\{2 m^8-8 m^6 t+m^4 \left(3 s^2+4 s t+12 t^2\right)-m^2 \left(s^3+2 s^2 t+8 s t^2+8 t^3\right)+t (s+t) \left(s^2+2 s t+2 t^2\right)\right\}.
\end{align}
The remaining terms represent the magnetic field corrections arising from the leading magnetic term $S_1(p)$ in Eq.~\ref{correctionPropagator}:

\begin{align}
      \nonumber  |\mathcal{M}|_{(1)}^2 &= \frac{\pi ^2 \alpha^2 ~(eB)^2 }{6 s^2 \left(m^2-t\right)^4 \left(-m^2+s+t\right)^4}\bigg\{s^2 \big(-106 m^{12}+m^{10} (145 s+412 t)-2 m^8 \left(87 s^2+169 s t+291 t^2\right) \\ \nonumber &+m^6 \left(113 s^3+146 s^2 t+126 s t^2+328 t^3\right)+m^4 \left(-28 s^4+48 s^3 t+270 s^2 t^2+200 s t^3-22 t^4\right)\\ \nonumber &-m^2 t \left(56 s^4+211 s^3 t+282 s^2 t^2+151 s t^3+36 t^4\right)+2 t (s+t) \left(3 s^4+11 s^3 t+14 s^2 t^2+6 s t^3+3 t^4\right)\big)\\ \nonumber & +4 s \cos 2 \theta  \big(10 m^{14}-4 m^{12} (13 s+15 t)+m^{10} \left(101 s^2+292 s t+150 t^2\right)-m^8 \left(86 s^3+495 s^2 t+675 s t^2+200 t^3\right)\\ \nonumber &+m^6 \left(47 s^4+376 s^3 t+960 s^2 t^2+820 s t^3+150 t^4\right)-m^4 \left(18 s^5+169 s^4 t+591 s^3 t^2+920 s^2 t^3+550 s t^4+60 t^5\right)\\ \nonumber &+m^2 (s+t) \left(4 s^5+36 s^4 t+145 s^3 t^2+253 s^2 t^3+182 s t^4+10 t^5\right)-s t (s+t) (2 s^4+16 s^3 t+43 s^2 t^2+54 s t^3 \\ \nonumber &+27 t^4)\big)+\cos 4 \theta  \big(160 m^{16}-160 m^{14} (3 s+8 t)+m^{12} \left(674 s^2+3520 s t+4480 t^2\right)-m^{10} (529 s^3+4484 s^2 t\\ \nonumber &+11040 s t^2+8960 t^3)+2 m^8 \left(115 s^4+1619 s^3 t+6165 s^2 t^2+9600 s t^3+5600 t^4\right)-m^6 \big(49 s^5+1362 s^4 t \\ \nonumber &+7702 s^3 t^2+17960 s^2 t^3+20000 s t^4  +8960 t^5\big)+2 m^4 (2 s^6+162 s^5 t+1379 s^4 t^2+4484 s^3 t^3+7315 s^2 t^4 \\ \nonumber &+6240 s t^5+2240 t^6)-m^2 t \left(40 s^6+533 s^5 t+2350 s^4 t^2+5137 s^3 t^3+6324 s^2 t^4+4320 s t^5+1280 t^6\right)\\  &+2 t (s+t) \big(s^6+21 s^5 t+108 s^4 t^2+254 s^3 t^3+327 s^2 t^4+240 s t^5+80 t^6\big)\big)\bigg\}
      \\
      \nonumber  |\mathcal{M}|_{(2)}^2 &=\frac{\pi ^2 \alpha^2 (eB)^2 \sin 2 \theta   }{3 s \left(m^2-t\right)^4 \left(-m^2+s+t\right)^4}\left(2 m^2-s-2 t\right) \sqrt{-m^4+2 m^2 t-t (s+t)}\big\{10 m^{10}-m^8 (47 s+40 t)\\ \nonumber &+4 m^6 \left(16 s^2+42 s t+15 t^2\right)-m^4 \left(39 s^3+182 s^2 t+222 s t^2+40 t^3\right)+2 m^2 (2 s^4+39 s^3 t+86 s^2 t^2+64 s t^3\\  &+5 t^4)-3 s t (s+t) \left(4 s^2+9 s t+9 t^2\right)\big\}
      \\
       \nonumber  |\mathcal{M}|_{(3)}^2 &= \frac{\pi ^2 \alpha^2 (eB)^2 \sin 2 \theta  \cos 2\theta}{3 s^2 \left(m^2-t\right)^4 \left(-m^2+s+t\right)^4}  \left(2 m^2-s-2 t\right) \sqrt{-m^4+2 m^2 t-t (s+t)} \big\{80 m^{12}-40 m^{10} (5 s+12 t)\\ \nonumber &+m^8 \left(237 s^2+1040 s t+1200 t^2\right)-2 m^6 \left(73 s^3+514 s^2 t+1080 s t^2+800 t^3\right)+m^4 (41 s^4+526 s^3 t+1662 s^2 t^2\\ \nonumber &+2240 s t^3+1200 t^4)-2 m^2 \left(2 s^5+65 s^4 t+307 s^3 t^2+594 s^2 t^3+580 s t^4+240 t^5\right)+t (s+t) (12 s^4+77 s^3 t\\  &+157 s^2 t^2+160 s t^3+80 t^4)\big\}
    \label{totalamplitudesquaredappendixlast}
\end{align}
The cross-section contribution from each of these terms is obtained by substituting them into Eq.~\ref{cross-section formula}. The terms $|\mathcal{M}|_{(0)}^2$ and $|\mathcal{M}|_{(1)}^2$ yield the leading-order vacuum cross section (Eq.~\ref{leadingorderCS}) and the magnetic field correction (Eq.~\ref{correctionCS}), respectively. The terms $|\mathcal{M}|_{(2)}^2$ and $|\mathcal{M}|_{(3)}^2$, which are proportional to $\sin 2\theta$ and $\sin 2\theta\cos 2\theta$ respectively, vanish identically upon integration over the Mandelstam variable $t$ in the cross-section formula. Consequently, the magnetic field correction to the total cross section retains only $\cos 2\theta$ and $\cos 4\theta$ dependence, as given in Eq.~\ref{correctionCS}.

\bibliographystyle{elsarticle-num}
\bibliography{references}	
\end{document}